\documentclass[12pt]{article}

\usepackage{amssymb}
\usepackage{luis}

\def\dsum{\mathop{\displaystyle \sum }}%
\def\stackunder#1#2{\mathrel{\mathop{#2}\limits_{#1}}}
\def\QATOP#1#2{{#1 \atop #2}}
\def\limfunc#1{\mathop{\rm #1}}%
\def\QTR#1#2{{\csname#1\endcsname #2}}
\def\binom#1#2{{#1 \choose #2}}

\newtheorem{theorem}{Theorem}[section]

\newtheorem{corollary}[theorem]{Corollary}
\newtheorem{lemma}[theorem]{Lemma}
\newtheorem{proposition}[theorem]{Proposition}
\newtheorem{definition}[theorem]{Definition}
\newtheorem{example}[theorem]{Example}

\newtheorem{remark}[theorem]{Remark}

\begin{document}

\thispagestyle{empty}

\vspace*{7cm}
\begin{center}
{\LARGE {\bf Marked Gibbs measures via \\[0.3cm] cluster expansion} \bigskip }

{\sc Yuri G.~Kondratiev}$^{1,2,3}$

{\sc Tobias Kuna}$^{1}$

{\sc Jos\'{e} L.~Silva}$^{2,4}$
\bigskip 

\begin{tabular}{l}
$^{1}$Inst.~Angewandte Math., Bonn Univ., D-53115 Bonn, Germany\\
$^{2}$BiBoS, Bielefeld Univ., D-33615 Bielefeld, Germany \\ 
$^{3}$Inst.~Math., 252601 Kiev, Ukraine\\ 
$^{4}$CCM, Univ.~Madeira, P-9000 Funchal, Portugal \\ 
\multicolumn{1}{r}{CCM$^{@}$-UMa {\bf 32}/98}
\end{tabular}
\end{center}

\renewcommand{\thefootnote}{}
\footnotetext{
$^{@}$http:/www.uma.pt/ccm/ccm.html}


\newpage
\renewcommand{\thefootnote}{\arabic{footnote}}
\setcounter{page}{1}

\title{
Marked Gibbs measures via cluster expansion}
\author{\textbf{Yuri G.~Kondratiev} \\
Inst.~Ang.~Math., Univ.~Bonn, D 53115 Bonn, Germany\\
BiBoS, Univ.~Bielefeld, D 33615 Bielefeld, Germany \\
Inst.~Math., NASU, 252601 Kiev, Ukraine\\ 
\and
\textbf{Tobias Kuna} \\
Inst.~Ang.~Math., Univ.~Bonn, D 53115 Bonn, Germany\\
\and
\textbf{Jos\'{e} L.~da Silva} \\
BiBoS, Univ.~Bielefeld, D 33615 Bielefeld, Germany \\
CCM, Univ.~Madeira, P 9000 Funchal, Portugal\\
}
\maketitle
%
%


\begin{abstract}
We give a sufficiently detailed account on the construction of marked Gibbs
measures in the high temperature and low fugacity regime. This is proved for
a wide class of underlying spaces and potentials such that stability and
integrability conditions are satisfied. That is, for state space we take a
locally compact separable metric space $X$ and a separable metric space $S$
for the mark space. This framework allowed us to cover several models of
classical and quantum statistical physics. Furthermore, we also show how to
extend the construction for more general spaces as e.g., separable standard
Borel spaces. The construction of the marked Gibbs measures is based on the
method of cluster expansion.
\end{abstract}

\newpage
\tableofcontents
\newpage

\section{Introduction\label{6eq1}}

The purpose of this paper is to give a detailed and comprehensive account on
the construction of marked Gibbs measures in the high temperature and low
fugacity regime for general underlying spaces using the method of cluster
expansion. Our motivation for this general framework is on the one hand
related to the examples in statistical physics we would like to cover, see
Examples~\ref{6eq112} - \ref{6eq126} below and also Subsection~\ref{6eq124}.
On the other hand, in recent papers \cite{AKR97}, \cite{AKR97a} (see also
lecture notes \cite{R98}) the authors put special emphasis in the
construction of differential geometry on the simple configuration space $%
\Gamma _X$ over a manifold $X$, i.e., 
\[
\Gamma _X:=\{\gamma \subset X\,|\,|\gamma \cap K|<\infty \;\mathrm{%
for\;any\;compact\;}K\subset X\},
\]
(cf.~(\ref{6eq90})) via a lifting of the geometry from the underlying
manifold $X$ (see as well \cite{KSS97} for an extension for compound Poisson
spaces). In \cite{AKR97b} the authors applied the aforementioned
differential geometry to construct representations of current algebras and
hence non-relativistic quantum field theories. This provides a scheme of
canonical quantizations which uses a Gibbs measure on the configuration
space as a ground state measure of the considered models. Having in mind the
study of quantum models with internal degrees of freedom we are interested
to extend the corresponding analysis to marked configurations and non flat
underlying spaces. It gives an additional motivation to develop analysis,
geometry, etc. on marked configuration spaces. In all applications mentioned
above marked Gibbs measures are playing a fundamental role. At present
moment any general results about the existence and uniqueness of marked
Gibbs measures are absent. The aim of our paper is to describe a
construction of such kind of measures in the case of general underlying and
marked space.

The results of this paper (which we will give an account below) are based on
the so-called cluster expansion method, see e.g., \cite{MM91}, \cite{P63}, 
\cite{R64}, and \cite{R69}, and we follow closely the scheme of
V.~A.~Malyshev and R.~A.~Minlos (cf.~\cite[Chap.~3 and 4]{MM91}), which the
authors realized for the configuration space over $\QTR{mathbb}{R}^d$. Let
us explain this more precisely. Let $X$ be a locally compact separable
metric space (the space describing the position of particles) and $S$ a
separable metric space (the mark space) describing some internal degrees of
freedom, e.g., spin, momentum, or different types of particles. We construct
a marked Poisson measure $\pi _\sigma ^\tau $ ($\sigma $ is an intensity
measure on $X$ and $\tau $ a transition kernel on $S$) over the marked
configuration space, i.e., 
\[
\Omega _X(S):=\{\omega =\{(x,s)\}\in \Gamma _{X\times S}|\{x\}=:\gamma
_\omega \in \Gamma _X\},
\]
via Kolmogorov's theorem, see Subsections~\ref{6eq6} and \ref{6eq92} below.
The desired measure $\mu $ on $\Omega _X(S)$ is obtained as a limit (in a
sense to be specified later) of a family of measures $\Pi _\Lambda ^{\sigma
^\tau ,\phi }$, cf.~Subsection~\ref{6eq122}. Here $\sigma ^\tau $ is the
measure defined on $(X\times S,\mathcal{B}(X\times S))$ by $\sigma ^\tau
(dx,ds)=\tau (x,ds)\sigma (dx)$, see (\ref{6eq125}) for details. For finite
volume $\Lambda \subset X$ (i.e., bounded Borel set) the measure $\Pi
_\Lambda ^{\sigma ^\tau ,\phi }$ is defined as a Gibbs type perturbation of
the marked Poisson measure $\pi _\sigma ^\tau $, i.e., 
\[
\Pi _\Lambda ^{\sigma ^\tau ,\phi }(\omega ,F):=\frac{1\!\!1_{\{Z_\Lambda
^{\sigma ^\tau ,\phi }<\infty \}}(\omega )}{Z_\Lambda ^{\sigma ^\tau ,\phi
}(\omega )}\int_\Omega 1\!\!1_F(\omega _{X\backslash \Lambda }\cup \omega
_\Lambda ^{\prime })e^{-E_\Lambda ^\phi (\omega _{X\backslash \Lambda }\cup
\omega _\Lambda ^{\prime })}\pi _\sigma ^\tau (d\omega ^{\prime }),
\]
(cf.~Definition~\ref{6eq113} in Section~\ref{6eq3}). It is well-known that $%
\Pi _\Lambda ^{\sigma ^\tau ,\phi }$ is a specification in the sense of 
\cite[Section~6]{P76} (see also \cite{P79} and \cite{P80}) for the given
pair potential $\phi $. Shortly speaking, a marked Gibbs measure is defined
as a probability measure which has as conditional expectation the
specification $\Pi _\Lambda ^{\sigma ^\tau ,\phi }$. The aforementioned
limit measure $\mu $ is locally absolutely continuous with respect to the
marked Poisson measure $\pi _\sigma ^\tau $ (cf.~Theorem~\ref{6eq58}). If we
assume additionally that the potential $\phi $ has finite range, then we
give a direct proof that the limit measure $\mu $ fulfils the DLR equation,
see Subsection~\ref{6eq130}, Theorem~\ref{6eq87}, and hence it is a Gibbs
measure. Let us mention that using further consequences of the cluster
expansion developed in \cite{Ku98} and the general results from \cite{KoKu98}
it is possible to show that the limit measure $\mu $ is a Gibbs measure for
a much wider class of potentials.

We would like to emphasize that the above results (specially the one of
Theorem~\ref{6eq58}) are strongly related with the procedure of cluster
expansion and the estimates obtained there. As usual, this procedure is
possible under some conditions on the potential $\phi $ and other parameters
of the system.

Thus the contents of Sections~\ref{6eq3}, \ref{6eq4}, and \ref{6eq5} has
been described. It remains to add that Section~\ref{6eq2} consists of the
necessary preliminaries for the further sections. Namely, we give a sketch
of the construction of the marked configuration space $\Omega _{X}(S)$ and
its measurable structure, (cf.~Subsection \ref{6eq6}) as well as the marked
Poisson measures $\pi _{\sigma }^{\tau }$, see Subsection~\ref{6eq92}. In
the remainder of Section~\ref{6eq2} we introduce some algebraic structures
in order to perform easier calculations and combinatorics involved in
cluster expansion. This is the contents of Subsection~\ref{6eq118} and \ref
{6eq35}. For the clarity of the presentation we moved some proofs to the
Appendix.

Finally, we would like to remark that all our results extends to underlying
spaces more general than we discuss in the main body of the work, namely,
separable standard Borel spaces. The necessary modifications are described
in Subsection~\ref{6eq76}. In a second paper, see \cite{Ku98}, we collect
further results for Gibbs measures in the high temperature regime.

\section{Marked configurations spaces \label{6eq2}}

In this section we describe the framework to be used in the rest of the
paper. Hence in Subsection~\ref{6eq6} we introduce the measurable structure
of the space on which the marked Gibbs measure will be defined, see Section~%
\ref{6eq3}. Let us mention that such measures are called \textit{states} in
statistical physics of continuous systems and in probability theory they are
known as \textit{marked point random fields}, cf.~e.g.~\cite{AGL78}, \cite
{GZ93}, \cite{K93}, and \cite{MM91}.

The marked Poisson measures are constructed in Subsection~\ref{6eq92}.
Finally, in Subsection~\ref{6eq118} (resp.~Subsection~\ref{6eq35}) we
introduce some facts from graph theory (resp.~*-calculus) which will
simplify our calculations later on, namely in Section~\ref{6eq4}.

Let $X$ be a locally compact separable metric space (which fulfils the
second axiom of countability, i.e., the topology is countably generated). It
describes the \textit{position space} of the particles. Denote by $\mathcal{B%
}(X)$ the Borel $\sigma $-algebra on $X$ and by $\mathcal{B}_c(X)$ the set
of all elements in $\mathcal{B}(X)$ which have compact closures (sets from $%
\mathcal{B}_c(X)$ we call finite volumes). Additionally, we suppose given a
complete separable metric space $S$. The corresponding Borel $\sigma $%
-algebra we denote by $\mathcal{B}(S)$. The elements of this space we call 
\textit{marks} (they can describe e.g., internal degrees of freedom).

\subsection{The marked configuration space over a manifold\label{6eq6}}

We briefly recall the basic definitions of the simple configuration space
over a manifold $X$ for the reader's convenience. The presentation is very
much based along the lines of the works by S.~Albeverio et al.~\cite{AKR97}.

The \textit{simple configuration space} $\Gamma :=\Gamma _X$ over the space $%
X$ is defined as the set of all locally finite subsets (configurations) in $%
X $: 
\begin{equation}
\Gamma _X:=\{\gamma \subset X\,|\,|\gamma \cap K|<\infty \;\mathrm{%
for\;any\;compact\;}K\subset X\}.  \label{6eq90}
\end{equation}
Here (and below) $|A|$ denotes the cardinality of a set $A$. For any $%
Y\subset X$ we define 
\[
\Gamma _Y:=\{\gamma \in \Gamma ||\gamma \cap (X\backslash Y)|=0\}. 
\]
In this paper we are interested in a bigger space of configurations, the
so-called marked configuration space, thus we proceed giving its abstract
definition. For concrete examples we refer to Subsection~\ref{6eq88}.

The \textit{marked configuration space} $\Omega _{X}(S):=\Omega _{X}:=\Omega 
$ is defined by 
\begin{equation}
\Omega :=\{\omega =\{(x,s)\}\in \Gamma _{X\times S}|\{x\}=:\gamma _{\omega
}\in \Gamma _{X},s\in S\}.  \label{6eq91}
\end{equation}
Equivalently $\Omega $ can be described as follows 
\[
\Omega :=\{\omega =(\gamma _{\omega },s)|\gamma _{\omega }\in \Gamma
_{X},s\in S^{\gamma _{\omega }}\}, 
\]
where $S^{\gamma _{\omega }}$ stands for the set of all maps $\gamma
_{\omega }\ni x\mapsto s_{x}\in S$. For any $Y\in \mathcal{B}(X)$ we define
in a similar way the space $\Omega _{Y}(S):=\Omega _{Y}$. We sometimes use
the shorthand $\omega _{Y}$ (resp.~$\gamma _{Y}$) for $\omega \cap (Y\times
S)$, $Y\subset X$ (resp.~$\gamma \cap Y$) and $\hat{x}:=(x,s_{x})\in X\times
S$.

In order to define a measurable structure on $\Omega $ we use the following
family of sets $\frak{I}$, the ``local'' sets 
\begin{equation}
\frak{I}:=\{B\in \mathcal{B}(X)\times \mathcal{B}(S)|\exists \Lambda \in 
\mathcal{B}_{c}(X)\;\mathrm{with}\;B\subset \Lambda \times S\}.
\label{6eq96}
\end{equation}

For any $A\in \frak{I}$ define the mapping $N_{A}:\Omega \rightarrow 
\QTR{mathbb}{N}_{0}$ by 
\[
N_{A}(\omega ):=|\omega \cap A|,\;\omega \in \Omega , 
\]
then 
\[
\mathcal{B}(\Omega ):=\sigma (\{N_{A}|A\in \frak{I}\}). 
\]
For any $Y\in \mathcal{B}(X)$ we define the following $\sigma $-algebra on $%
\Omega $%
\[
\mathcal{B}_{Y}(\Omega ):=\sigma (\{N_{A}|A\in \frak{I},\;A\subset Y\times
S\}). 
\]

For any $Y\in \mathcal{\ B}(X)$ the $\sigma $-algebra $\mathcal{B}_Y(\Omega
) $ is isomorphic to $\mathcal{B}(\Omega _Y)$. The ``filtration'' $(\mathcal{%
B}_\Lambda (\Omega ))_{\Lambda \in \mathcal{B}_c(X)}$ is one of the basic
structures in the definition of the Gibbs measures, see Section~\ref{6eq3}
and \ref{6eq5}. Moreover, if $Y_1,Y_2\in \mathcal{B}(X)$ such that $Y_1\cap
Y_2\neq \emptyset $, then $\Omega _{Y_1\sqcup Y_2}$ is isomorphic to $\Omega
_{Y_1}\times \Omega _{Y_2}$.

Finally we want to give another useful description of the marked
configuration space $\Omega $. For any $n\in \QTR{mathbb}{N}_{0}$ and any $%
Y\in \mathcal{B}(X)$ we define the $n$-\textit{point configuration space} $%
\Omega _{Y}^{(n)}$ as a subset of $\Omega _{Y}$ by 
\[
\Omega _{Y}^{(n)}:=\Omega _{Y}^{(n)}(S):=\{\omega \in \Omega _{Y}||\omega
|=n\},\;\Omega _{Y}^{(0)}:=\{\emptyset \}, 
\]
and denote the corresponding $\sigma $-algebra by $\mathcal{B}(\Omega
_{Y}^{(n)})$.

There is a bijection 
\begin{equation}
(\widetilde{Y\times S})^{n}/S_{n}\rightarrow \Omega _{Y}^{(n)},\;n\in 
\QTR{mathbb}{N},\;Y\in \mathcal{B}(X),  \label{6eq102}
\end{equation}
where 
\[
(\widetilde{Y\times S})^{n}:=\{((x_{1},s_{x_{1}}),\ldots
,(x_{n},s_{x_{n}}))|x_{i}\in Y,s_{x_{i}}\in S,x_{i}\neq x_{j},\;\mathrm{for}%
\;i\neq j\}, 
\]
and $S_{n}$ denotes the permutation group over $\{1,\ldots ,n\}.$ Since this
bijection is measurable in both directions the natural $\sigma $-algebra on $%
(\widetilde{Y\times S})^{n}/S_{n}$ is isomorphic to $\mathcal{B}(\Omega
_{Y}^{(n)})$.

One can reconstruct $\Omega $ from the sets $\Omega _{\Lambda }^{(n)}$ using
the following scheme. First notice that we can write for any $\Lambda \in 
\mathcal{B}_{c}(X)$%
\[
\Omega _{\Lambda }=\bigsqcup_{n=0}^{\infty }\Omega _{\Lambda }^{(n)}, 
\]
hence the $\sigma $-algebra $\mathcal{B}(\Omega _{\Lambda })$ is the
disjoint union of the $\sigma $-algebras $\mathcal{B}(\Omega _{\Lambda
}^{(n)})$.

For any $\Lambda _{1},\Lambda _{2}\in \mathcal{B}_{c}(X)$ with $\Lambda
_{1}\subset \Lambda _{2}$ there are natural maps

\[
p_{\Lambda _{2},\Lambda _{1}}:\Omega _{\Lambda _{2}}\longrightarrow \Omega
_{\Lambda _{1}}, 
\]
\[
p_{\Lambda _{1}}:\Omega \longrightarrow \Omega _{\Lambda _{1}} 
\]
defined by $p_{\Lambda _{2},\Lambda _{1}}(\omega ):=\omega _{\Lambda _{1}}$, 
$\omega \in \Omega _{\Lambda _{2}}$ (resp.~$p_{\Lambda _{1}}(\omega )=\omega
_{\Lambda _{1}}$, $\omega \in \Omega $). It can be shown that $(\Omega ,%
\mathcal{B}(\Omega ))$ coincides with the projective limit of the measurable
spaces $(\Omega _{\Lambda },\mathcal{B}(\Omega _{\Lambda }))$, $\Lambda \in 
\mathcal{B}_{c}(X)$.

Finally, we would like to introduce one more subspace of $\Omega $ which
plays a fundamental role in our calculations below, the \textit{finite
configuration space} $\Omega _{X,fin}:=\Omega _{fin}$. It is defined by 
\[
\Omega _{fin}:=\{\omega \in \Omega |\,|\omega |<\infty \}. 
\]
The finite configuration space $\Omega _{fin}$ has the following useful
representation in terms of the $n$-point configuration spaces 
\begin{equation}
\Omega _{fin}=\bigsqcup_{n=0}^\infty \Omega _X^{(n)}.  \label{6eq89}
\end{equation}
Analogously for $\Omega _{Y,fin}$, $Y\in \mathcal{B}(X)$. The space $\Omega
_{fin}$ (resp.~$\Omega _{Y,fin}$) is equipped with the $\sigma $-algebra $%
\mathcal{B}(\Omega _{fin})$ (resp.~$\mathcal{B}(\Omega _{Y,fin})$) of the
disjoint unions of measurable spaces $(\Omega _X^{(n)},\mathcal{B}(\Omega
_X^{(n)}))$ (resp.~$(\Omega _Y^{(n)},\mathcal{B}(\Omega _Y^{(n)}))$.

\subsection{Marked Poisson measures\label{6eq92}}

For constructing the marked Poisson measure on $\Omega $ we need, first of
all, to fix an intensity measure $\sigma $ on the underlying space $X$.
Thus, let us assume that $\sigma $ is a non-atomic Radon measure on $X$.
Additionally, we define a kernel $\tau :X\times \mathcal{B}(S)\rightarrow 
\QTR{mathbb}{R}$, i.e., $\forall x\in X$ $\tau (x,\cdot )$ is a finite
measure on $(S,\mathcal{B}(S))$ and $\tau (\cdot ,A)$ is $\mathcal{B}(X)$%
-measurable for all $A\in \mathcal{B}(S)$. Moreover we assume that the
following condition is fulfilled for any $\Lambda \in \mathcal{B}_c(X)$%
\begin{equation}
\int_\Lambda \tau (x,S)\sigma (dx)<\infty .  \label{6eq57}
\end{equation}
This condition will be essential in the estimates later on (cf.~proof of
Proposition \ref{6eq74}).

In the product space $X\times S$ we define a $\sigma $-finite measure $%
\sigma ^{\tau }$ by 
\[
\sigma ^{\tau }(dx,ds):=\tau (x,ds)\sigma (dx), 
\]
that means for $A\times B\in \mathcal{B}(X\times S)$%
\begin{equation}
\sigma ^{\tau }(A\times B)=\int_{A}\tau (x,B)\sigma (dx),  \label{6eq125}
\end{equation}
which is a non-atomic Radon measure.

For any $Y\in \mathcal{B}(X)$ and $n\in \QTR{mathbb}{N}$ the product measure 
$\sigma ^{\tau \otimes n}$ can be considered as a measure on $(\widetilde{%
Y\times S})^{n}$, cf.~Lemma~\ref{6eq120} in the Appendix. Let 
\[
\sigma _{n}^{\tau }\upharpoonright \Omega _{Y}^{(n)}:=\sigma ^{\tau \otimes
n}\circ (\mathrm{sym}_{Y}^{n})^{-1}, 
\]
be the corresponding measure on $\Omega _{Y}^{(n)}$, where 
\[
\mathrm{sym}_{Y}^{n}:(\widetilde{Y\times S})^{n}\rightarrow \Omega
_{Y}^{(n)}, 
\]
given by 
\[
\mathrm{sym}_{Y}^{n}((\hat{x}_{1},\ldots ,\hat{x}_{n})):=\{\hat{x}%
_{1},\ldots ,\hat{x}_{n}\}\in \Omega _{Y}^{(n)}. 
\]

Then we consider the so-called \textit{Lebesgue-Poisson measure} $\nu
_{z\sigma ^\tau }$ on $\mathcal{B}(\Omega _{fin})$, which coincides on each $%
\Omega _X^{(n)}$ with the measure $\frac{z^n}{n!}\sigma _n^\tau
\upharpoonright \Omega _X^{(n)}$, as follows 
\begin{equation}
\nu _{z\sigma ^\tau }:=\sum_{n=0}^\infty \frac{z^n}{n!}(\sigma _n^\tau
\upharpoonright \Omega _X^{(n)}),  \label{6eq93}
\end{equation}
and $\sigma _0^\tau (\emptyset ):=1$. As $a$ result $\nu _{z\sigma ^\tau }$
is $\sigma $-finite. $z>0$ is the so called activity parameter.

Considered as a measure on $\Omega _{\Lambda }$, $\Lambda \in \mathcal{B}%
_{c}(X)$, the measure $\nu _{z\sigma ^{\tau }}$ is finite with $\nu
_{z\sigma ^{\tau }}(\Omega _{\Lambda })=e^{z\sigma ^{\tau }(\Lambda \times
S)}$. Therefore, we can define a probability measure $\pi _{z\sigma }^{\tau
,\Lambda }$ on $\Omega _{ \Lambda }$ putting 
\[
\pi _{z\sigma }^{\tau ,\Lambda }:=e^{-z\sigma ^{\tau }(\Lambda \times S)}\nu
_{z\sigma ^{\tau }}. 
\]
The measure $\pi _{z\sigma }^{\tau ,\Lambda }$ has the following property 
\[
\pi _{z\sigma }^{\tau ,\Lambda }(\Omega _{\Lambda }^{(n)})=\frac{z^{n}}{n!}%
(\sigma ^{\tau }(\Lambda \times S))^{n}e^{-z\sigma ^{\tau }(\Lambda \times
S)}, 
\]
which gives the probability of the occurrence of exactly $n$ points of the
marked Poisson process (with arbitrary values of marks) inside of the volume 
$\Lambda $.

In order to obtain the existence of a unique probability measure $\pi
_{z\sigma }^\tau $ on $(\Omega ,\mathcal{B}(\Omega ))$ such that 
\[
\pi _{z\sigma }^{\tau ,\Lambda }=\pi _{z\sigma }^\tau \circ p_\Lambda
^{-1},\;\Lambda \in \mathcal{B}_c(X), 
\]
we notice that the family $\{\pi _{z\sigma }^{\tau ,\Lambda }|\Lambda \in 
\mathcal{B}_c(X)\}$ is consistent, i.e., 
\[
\pi _{z\sigma }^{\tau ,\Lambda _2}\circ p_{\Lambda _2,\Lambda _1}^{-1}=\pi
_{z\sigma }^{\tau ,\Lambda _1},\;\Lambda _1,\Lambda _2\in \mathcal{B}%
_c(X),\Lambda _1\subset \Lambda _2, 
\]
and thus, by a version of Kolmogorov's theorem for the projective limit
space $\Omega $ (cf.~\cite[Chap.~V Theorem~3.2]{P67} or Theorem~\ref{6eq100}
below) any such family determines uniquely a measure $\pi _{z\sigma }^\tau $
on $\mathcal{B}(\Omega )$ such that $\pi _{z\sigma }^{\tau ,\Lambda }=\pi
_{z\sigma }^\tau \circ p_\Lambda ^{-1}$. The measure $\pi _{z\sigma }^\tau $
is called \textit{marked Poisson measure}.

\subsection{Basic concepts in graph theory\label{6eq118}}

Now we are going to introduce some standard concepts of graph theory, see
e.g.,~\cite{O67} for more details.

Let $X$ be a non empty set. A \textit{partition} of $X$ is a family of non
empty subsets $(X_i)_{i\in I}$ of $X,$ called parts, such that $X_i\cap
X_j=\emptyset $ for $i\neq j$ and $\bigcup_iX_i=X$. The set of all
partitions of $X$ where all parts are non-empty is denoted by $\frak{P}(X)$
and by $\frak{P}^n(X)$ we denote the subset of partitions of $\frak{P}(X)$
consisting of $n$ parts. $\frak{P}_\emptyset ^n(X)$ stands for the set of
all partitions of $n$ parts which might be empty.

We now give the notion of a graph as well as some of its properties. We note
here and henceforth that the graphs under consideration are undirected, see 
\cite[Chap.~1]{O67} for this notion.

\begin{definition}
\label{6eq108}

\begin{enumerate}
\item  A graph $G:=G(X)$ is a subset of 
\[
\{\{x,y\}\subset X|x\neq y\}. 
\]
One calls $\{x,y\}\in G$ the edges of the graph and $V(G):=X$ the vertices
of the graph. The collection of all such graphs on $X$ is denoted by $\frak{G%
}(X).$

\item  Given two graphs $G_1\in \frak{G}(X_1)$, $G_2\in \frak{G}(X_2)$ with $%
G_1\cap G_2=\emptyset $, their sum graph is the graph given by 
\[
G_1\sqcup G_2=\{\{x,y\}\subset X_1\cup X_2|\{x,y\}\in G_1\sqcup G_2\}. 
\]
If $G_1$ and $G_2$ have no common vertices, then the sum graph is denoted by 
$G_1\oplus G_2$. This procedure extends to an arbitrary family $\{G_i\}$ of
graphs.

\item  A graph $G$ is called connected iff any pair of vertices is
connected. The set of all connected graphs in $X$ is denoted by $\frak{G}%
^c(X)$. We assume that the single point is a connected graph.
\end{enumerate}
\end{definition}

\begin{proposition}
\label{6eq109}(cf.~\cite[Theorem 2.2.1]{O67}) Let $G\in \frak{G}$ be given.
Then $G$ decomposes uniquely into a disjoint sum $\oplus _iG_i$ of its
connected components.
\end{proposition}

\begin{definition}
A connected graph $G$ is called a tree iff it has no loops. The set of all
trees on $X$ is denoted by $\frak{T}(X)$.
\end{definition}

\begin{proposition}
\label{6eq110}(cf.~\cite[Theorem 4.1.3]{O67}) The number of different trees
which can be constructed on $n$ given vertices is $n^{n-2}$.
\end{proposition}

\begin{remark}
\label{6eq131}

\begin{enumerate}
\item  Since for any $n\geq 0$ we have 
\[
\sqrt{2\pi n}n^ne^{-n}\leq n!\leq \sqrt{2\pi n}n^ne^{-n}\exp \left( \frac
1{12(n-1)}\right) ,
\]
it is not hard to see that $n^{n-2}<e^nn!$.

\item  We use the shorthand $[n]$ for $\{1,\ldots ,n\}$ and thus the symbol $%
\frak{T}([n])$ denotes the trees in $\{1,\ldots ,n\}$.
\end{enumerate}
\end{remark}

\subsection{*-calculus\label{6eq35}}

In this subsection we point out an algebraic structure (see, e.g.\cite{R64}, 
\cite{R69}, and \cite{MM91},) which turns out to simplify our notation and
calculations later on. It will be very interesting to clarify more the
related analytic and algebraic structure of this calculus.

Let $\mathcal{A}$ be the set of all measurable (complex-valued) functions $%
\psi $ on $\Omega _{fin}$, i.e., 
\[
\mathcal{A}:=\{\psi :\Omega _{fin}\rightarrow \QTR{mathbb}{C},\psi \;\mathrm{%
is\;\mathcal{B}}(\Omega _{fin})\mathrm{-measurable}\}. 
\]

In $\mathcal{A}$ we introduce the following operation: for any $\psi _1,\psi
_2\in \mathcal{A}$ and $\omega \in \Omega _{fin}$ we define $\psi _1*\psi _2$
by 
\[
(\psi _1*\psi _2)(\omega ):=\sum_{(\omega _1,\omega _2)\in \frak{P}%
_\emptyset ^2(\omega )}\psi _1(\omega _1)\psi _2(\omega _2),\quad \omega \in
\Omega _{fin}, 
\]
which is $\mathcal{B}(\Omega _{fin})$-measurable because the restriction to $%
\Omega _X^{(n)}$ is of the form 
\[
(\psi _1*\psi _2)(\{\hat{x}_1,\ldots ,\hat{x}_n\})=\sum_{(I,J)\in \frak{P}%
_\emptyset ^2([n])}\psi _1(\{\hat{x}_i|i\in I\})\psi _2(\{\hat{x}_j|j\in
J\}). 
\]
The set $\mathcal{A}$ equipped with $*$ and the natural vector space
structure forms a commutative algebra with unit element 
\begin{equation}
1^{*}(\omega )=\left\{ 
\begin{array}{cc}
1, & \omega =\emptyset \\ 
0, & \omega \neq \emptyset
\end{array}
\right. .  \label{6eq79}
\end{equation}

Notice that for any $\psi _1,\ldots \psi _n\in \mathcal{A}$ we have 
\begin{equation}
(\psi _1*\ldots *\psi _n)(\omega )=\sum_{(\omega _1,\ldots ,\omega _n)\in 
\frak{P}_\emptyset ^n(\omega )}\psi _1(\omega _1)\ldots \psi _n(\omega
_n),\;\,\,\omega \in \Omega _{fin}.  \label{6eq36}
\end{equation}

Let us define $\mathcal{A}_0$ as a subset of $\mathcal{A}$ by 
\[
\mathcal{A}_0:=\{\psi \in \mathcal{A}|\psi (\emptyset )=0\}. 
\]
For any $\psi \in \mathcal{A}$ and $\varphi \in \mathcal{A}_0$ we have 
\begin{eqnarray*}
(\psi *\varphi )(\emptyset ) &=&\sum_{(\omega _1,\omega _2)\in \frak{P}%
_\emptyset ^2(\emptyset )}\psi (\omega _1)\varphi (\omega _2) \\
&=&\psi (\emptyset )\varphi (\emptyset )=0,
\end{eqnarray*}
thus it follows that $\mathcal{A}_0$ is an ideal in $\mathcal{A}$.

Let us introduce the mapping $\exp ^{*}:\mathcal{A}_0\rightarrow 1^{*}+%
\mathcal{A}_0$ defined by 
\begin{equation}
\exp ^{*}\psi :=\sum_{n=0}^\infty \frac 1{n!}\psi ^{*n}=1^{*}+\psi +\frac
1{2!}\psi ^{*2}+\ldots +\frac 1{n!}\psi ^{*n}+\ldots .  \label{6eq48}
\end{equation}
It follows from (\ref{6eq36}) that for any $\psi \in \mathcal{A}_0$%
\begin{eqnarray*}
(\exp ^{*}\psi )(\emptyset ) &=&1^{*}, \\
(\exp ^{*}\psi )(\omega ) &=&\sum_{n=0}^\infty \frac 1{n!}\sum_{(\omega
_1,\ldots ,\omega _n)\in \frak{P}_\emptyset ^n(\omega )}\psi (\omega
_1)\ldots \psi (\omega _n),\;\omega \in \Omega _{fin}\backslash \{\emptyset
\}.
\end{eqnarray*}

Moreover, if we define the mapping $\ln ^{*}:1^{*}+\mathcal{A}_0\rightarrow 
\mathcal{A}_0$ by 
\[
\ln ^{*}(1^{*}+\psi ):=\sum_{n=1}^\infty \frac{(-1)^{n-1}}n\psi ^{*n}, 
\]
then $\exp ^{*}$ and $\ln ^{*}$ are inverse one each other.

For simplicity, in what follows we introduce some notation: $\{\hat{x}\}_1^n$
denotes $\{\hat{x}_1,\ldots ,\hat{x}_n\}$ and $\sigma ^\tau (d\hat{x})_1^n$
denotes $\sigma ^\tau (dx_1,ds_{x_1})\dots \sigma ^\tau (dx_n,ds_{x_n})$
(analogous for $\nu _{z\sigma ^\tau }(d\omega )_1^n$).

Next we prove some lemmas which will be useful later on.

\begin{lemma}
\label{6eq38}Let $F,\psi _1,\ldots ,\psi _n$ be $\mathcal{B}(\Omega _{fin})$%
-measurable functions. Then the following equality holds: 
\begin{eqnarray}
\int_{\Omega _{fin}}F(\omega )(\psi _1*\ldots *\psi _n)(\omega )\nu
_{z\sigma ^\tau }(d\omega )  \label{6eq39} \\
=\int_{\Omega _{fin}}\dots \int_{\Omega _{fin}}F(\omega _1\cup \dots \cup
\omega _n)\psi _1(\omega _1)\dots \psi _n(\omega _2)\nu _{z\sigma ^\tau
}(d\omega )_1^n,  \nonumber
\end{eqnarray}
whenever all functions are positive or one side make sense for the modulus
of the functions.
\end{lemma}

\noindent \textbf{Proof.} Let $F,\psi _1,\ldots \psi _p$ be as
aforementioned. Then the definition of $\nu _{z\sigma ^\tau }$ on the right
hand side of (\ref{6eq39}) gives 
\begin{eqnarray*}
&&\sum_{n_1,\ldots ,n_p=0}^\infty \frac{z^{n_1+\ldots +n_p}}{n_1!\dots n_p!}%
\int_{X^{n_1}}\int_{S^{_{n_1}}}\dots \int_{X^{n_p}}\int_{S^{n_p}}F(\{\hat{x}%
\}_1^{n_1+\ldots +n_p}) \\
&&\times \psi _1(\{\hat{x}\}_1^{n_1})\dots \psi _p(\{\hat{x}\}_{n_1+\dots
+n_{p-1}+1}^{n_1+\dots +n_p})\sigma ^\tau (d\hat{x})_1^{n_1+\ldots +n_p} \\
&=&\sum_{n=0}^\infty \frac{z^n}{n!}\sum_{n_1+\ldots +n_p=n}\frac{n!}{%
n_1!\dots n_p!}\int_{X^n}\int_{S^n}F(\{\hat{x}\}_1^n) \\
&&\times \psi _1(\{\hat{x}\}_1^{n_1})\dots \psi _p(\{\hat{x}\}_{n_1+\dots
+n_{p-1}+1}^n)\sigma ^\tau (dx,ds)_1^n.
\end{eqnarray*}
Then interchanging the second sum with the integrals and using the
definition of $\nu _{z\sigma ^\tau }$ we derive the desired result.\hfill $%
\blacksquare $

\begin{corollary}
\label{6eq40}For any $Y\in \mathcal{B}(X)$ and $\psi \in \mathcal{A}$ such
that either $\psi $ positive or $\psi \in L^1(\Omega _{Y,fin},\nu _{z\sigma
^\tau })$ the following equality holds 
\begin{equation}
\int_{\Omega _{Y,fin}}(\exp ^{*}\psi )(\omega )\nu _{z\sigma ^\tau }(d\omega
)=\exp \left( \int_{\Omega _{Y,fin}}\psi (\omega )\nu _{z\sigma ^\tau
}(d\omega )\right) .  \label{6eq42}
\end{equation}
\end{corollary}

\begin{lemma}
\label{6eq41}Let $\psi \in \mathcal{A}$ and $\Lambda ,\Lambda ^{\prime }\in 
\mathcal{B}_c(X)$ be given such that $\Lambda ^{\prime }\subset \Lambda $,
suppose that $\psi \in L^1(\Omega _\Lambda ,\nu _{z\sigma ^\tau })$. Then
the following equality holds 
\begin{eqnarray}
&&\int_{\Omega _{\Lambda \backslash \Lambda ^{\prime }}}(\exp ^{*}\psi
)(\omega \cup \omega ^{\prime })\nu _{z\sigma ^\tau }(d\omega )
\label{6eq43} \\
&=&\exp \left( \int_{\Omega _{\Lambda \backslash \Lambda ^{\prime
}}}\!\!\!\!\!\!\psi (\omega )\nu _{z\sigma ^\tau }(d\omega )\right) \exp
^{*}\left( \int_{\Omega _{\Lambda \backslash \Lambda ^{\prime }}}\!\!\!\
\!\!1\!\!1_{\Omega _{fin}\backslash \{\emptyset \}}(\cdot )\psi (\cdot \cup
\omega )\nu _{z\sigma ^\tau }(d\omega )\right) (\omega ^{\prime }), 
\nonumber
\end{eqnarray}
for $\nu _{z\sigma ^\tau }$-a.e.~$\omega ^{\prime }\in \Omega _{\Lambda
^{\prime }}$.
\end{lemma}

The details of the proof are given in Appendix~\ref{6eq73}.

\begin{definition}
\label{6eq160}Let $\psi \in \mathcal{A}$. We define a ``differential''
operator $D$ by setting for $\omega ,\omega ^{\prime }\in \Omega _{fin}$ 
\begin{equation}
(D_{\omega ^{\prime }}\psi )(\omega ):=\psi (\omega \cup \omega ^{\prime
}),\;\mathrm{if\;}\gamma _\omega \cap \gamma _{\omega ^{\prime }}=\emptyset ,
\label{6eq119}
\end{equation}
and $(D_{\omega ^{\prime }}\psi )(\omega )=0$ otherwise.
\end{definition}

\begin{remark}
Let us mention that the operator $D$ is related with the Poissonian gradient 
$\nabla ^P$ (see e.g., \cite{KSS96} and \cite{NV95}) by 
\[
(\nabla ^P\psi )(\omega ,\hat{x})=(D_{\{\hat{x}\}}\psi )(\omega )-\psi
(\omega ). 
\]
\end{remark}

Finally, we state some properties of the operator $D$, which can be easily
checked using the Definition~\ref{6eq160}.

\begin{proposition}
\label{6eqT1}Let $\psi ,\psi _1,\psi _2\in \mathcal{A}$, $\omega \in \Omega
_{fin}$, $\hat{x},\hat{y}\in X\times S$, with $x\neq y$ and $x\notin \gamma
_\omega $ then

\begin{enumerate}
\item  $D_{\{\hat{x}\}}D_{\{\hat{y}\}}=D_{\{\hat{y}\}}D_{\{\hat{x}\}}$.

\item  \label{6eq46}$\left[ D_{\{\hat{x}\}}(\psi _1*\psi _2)\right] \left(
\omega \right) =\left[ (D_{\{\hat{x}\}}\psi _1)*\psi _2+\psi _1*(D_{\{\hat{x}%
\}}\psi _2)\right] (\omega )$.

\item  \label{6eq47}$\left[ D_{\{\hat{x}\}}\exp ^{*}\psi \right] (\omega
)=\left[ (\exp ^{*}\psi )*(D_{\{\hat{x}\}}\psi )\right] (\omega )$.
\end{enumerate}
\end{proposition}

\section{Marked Gibbs measures\label{6eq3}}

In the previous section we introduced the probability measure $\pi _\sigma
^\tau $ on $(\Omega ,\mathcal{B}(\Omega ))$, the so-called marked Poisson
measure, cf.~Subsection \ref{6eq92}. Now we will describe a more wide class
of probability measures on $(\Omega ,\mathcal{B}(\Omega )),$ namely, the 
\textit{marked Gibbs measures}. In Subsection~\ref{6eq88} we state various
examples and the associated marked Gibbs measures will be considered in
Subsection~\ref{6eq124}.

\subsection{Specifications, Gibbs measures, and global conditions}

A symmetric measurable function $\phi :(X\times S)\times (X\times
S)\rightarrow \QTR{mathbb}{R}\cup \{+\infty \}$ is called a \textit{pair
potential}. For a given pair potential we define the \textit{energy
functional} $E^\phi :\Omega _{fin}\rightarrow \QTR{mathbb}{R}\cup \{+\infty
\}$ by 
\begin{equation}
E^\phi (\omega ):=\dsum\limits_{\{\hat{x},\hat{y}\}\subset \omega }\phi (%
\hat{x},\hat{y}),  \label{6eq18}
\end{equation}
with $E^\phi (\emptyset ):=0$.

Let $\omega \in \Omega _{fin}$ and $\zeta \in \Omega $ be given, then the 
\textit{interaction energy }between $\omega $ and $\zeta $ is given by 
\begin{equation}
W^\phi (\omega ,\zeta ):=\left\{ 
\begin{array}{cl}
\dsum\limits_{\hat{x}\in \omega ,\hat{y}\in \zeta }\phi (\hat{x},\hat{y}), & 
\mathrm{if\;}\dsum\limits_{\hat{x}\in \omega ,\hat{y}\in \zeta }|\phi (\hat{x%
},\hat{y})|<\infty \\ 
&  \\ 
+\infty , & \mathrm{otherwise}
\end{array}
\right. .  \label{6eq78}
\end{equation}
For any $\Lambda \in \mathcal{B}_c(X)$ the \textit{conditional energy} $%
E_\Lambda ^\phi :\Omega \rightarrow \QTR{mathbb}{R}\cup \{+\infty \}$ is
defined by 
\[
E_\Lambda ^\Phi (\omega )=E^\phi (\omega _\Lambda )+W^\phi (\omega _\Lambda
,\omega _{X\backslash \Lambda }). 
\]

Notice that the energy $E^{\phi }$ may be expressed for any $\omega ,\omega
^{\prime }\in \Omega _{fin}\backslash \{\emptyset \}$ such that $\gamma
_{\omega }\cap \gamma _{\omega ^{\prime }}=\emptyset $ as 
\begin{equation}
E^{\phi }(\omega \cup \omega ^{\prime })=E^{\phi }(\omega )+E^{\phi }(\omega
^{\prime })+W(\omega ,\omega ^{\prime }).  \label{6eq77}
\end{equation}

Now we can define grand canonical marked Gibbs measures.

\begin{definition}
\label{6eq113}For any $\Lambda \in \mathcal{B}_c(X)$ the marked
specification $\Pi _\Lambda ^{\sigma ^\tau ,\phi }$ is defined for any $%
\omega \in \Omega $, $F\in \mathcal{B}(\Omega )$ by (see \cite{P76}) 
\begin{eqnarray}
&&\Pi _\Lambda ^{\sigma ^\tau ,\phi }(\omega ,F)\mbox{$:=$}1\!\!1_{\{\tilde{Z%
}_\Lambda ^{\sigma ^\tau ,\phi }<\infty \}}(\omega )[\tilde{Z}_\Lambda
^{\sigma ^\tau ,\phi }(\omega )]^{-1}\int_{\Omega _\Lambda }1\!\!1_F(\omega
_{X\backslash \Lambda }\cup \omega ^{\prime }) \\
&&\times \exp [-\beta E_\Lambda ^\phi (\omega _{X\backslash \Lambda }\cup
\omega ^{\prime })]\nu _{z\sigma ^\tau }(d\omega ^{\prime }),  \label{6eq19}
\end{eqnarray}
where $\beta >0$ is the inverse temperature. $\tilde{Z}_\Lambda ^{\sigma
^\tau ,\phi }$ is called partition function: 
\begin{equation}
\tilde{Z}_\Lambda ^{\sigma ^\tau ,\phi }(\omega ):=\int_{\Omega _\Lambda
}\exp [-\beta E_\Lambda ^\phi (\omega _{X\backslash \Lambda }\cup \omega
^{\prime })]\nu _{z\sigma ^\tau }(d\omega ^{\prime }).  \label{6eq20}
\end{equation}

A probability measure $\mu $ on $(\Omega ,\mathcal{B}(\Omega ))$ is called a
grand canonical marked Gibbs measure with interaction potential $\phi $ iff 
\begin{equation}
\mu \Pi _\Lambda ^{\sigma ^\tau ,\phi }=\mu ,\;\mathrm{for\;all}\;\Lambda
\in \mathcal{B}_c(X),  \label{6eq21}
\end{equation}
where for any $F\in \mathcal{B}(\Omega )$ the measure $\mu \Pi _\Lambda
^{\sigma ^\tau ,\phi }$ is defined by 
\[
(\mu \Pi _\Lambda ^{\sigma ^\tau ,\phi })(F):=\int_\Omega \Pi _\Lambda
^{\sigma ^\tau ,\phi }(\omega ,F)d\mu (\omega ), 
\]
and (\ref{6eq21}) above are called Dobrushin-Landford-Ruelle (DLR) equations.

Let $\mathcal{G}_{gc}(\sigma ^\tau ,\phi )$ denote the set of all such
probability measures $\mu $.
\end{definition}

\begin{remark}
\label{6eq117}

\begin{enumerate}
\item  It is well-known that $\{\Pi _\Lambda ^{\sigma ^\tau ,\phi
}\}_{\Lambda \in \mathcal{B}_c(X)}$ is a $\{\mathcal{B}_{X\backslash \Lambda
}(\Gamma )\}_{\Lambda \in \mathcal{B}_c(X)}$-specification in the following
sense (see e.g., \cite{Fo75}, \cite{P76}, \cite{P79}), for all $\Lambda
,\Lambda ^{\prime }\in \mathcal{B}_c(X).$

\begin{enumerate}
\item[(S1)]  $\Pi _\Lambda ^{\sigma ^\tau ,\phi }(\omega ,\Omega )\in
\{0,1\} $ for all $\omega \in \Omega $.

\item[(S2)]  $\Pi _\Lambda ^{\sigma ^\tau ,\phi }(\cdot ,Y)$ is $\mathcal{B}%
_{X\backslash \Lambda }(\Omega )$-measurable for all $Y\in \mathcal{B}%
(\Omega )$.

\item[(S3)]  $\Pi _\Lambda ^{\sigma ^\tau ,\phi }(\cdot ,Y\cap Y^{\prime
})=1\!\!1_{Y^{\prime }}\Pi _\Lambda ^{\sigma ^\tau ,\phi }(\cdot ,Y)$ for
all $Y\in \mathcal{B}(\Omega )$, $Y^{\prime }\in \mathcal{B}_{X\backslash
\Lambda }(\Omega )$.

\item[(S4)]  $\Pi _{\Lambda ^{\prime }}^{\sigma ^\tau ,\phi }=\Pi _{\Lambda
^{\prime }}^{\sigma ^\tau ,\phi }\Pi _\Lambda ^{\sigma ^\tau ,\phi }$ if $%
\Lambda \subset \Lambda ^{\prime }$. Here for any $\omega \in \Omega $, $%
Y\in \mathcal{B}(\Omega )$%
\[
(\Pi _{\Lambda ^{\prime }}^{\sigma ^\tau ,\phi }\Pi _\Lambda ^{\sigma ^\tau
,\phi })(\omega ,Y):=\int_\Omega \Pi _\Lambda ^{\sigma ^\tau ,\phi }(\omega
^{\prime },Y)\Pi _{\Lambda ^{\prime }}^{\sigma ^\tau ,\phi }(\omega ,d\omega
^{\prime }). 
\]
\end{enumerate}

\item  It can be easily shown that because of (\ref{6eq93}) for all $\Lambda
\in \mathcal{B}_c(X)$, $\omega \in \Omega $, $F\in \mathcal{B}(\Omega )$%
\begin{eqnarray*}
&&\Pi _\Lambda ^{\sigma ^\tau ,\phi }(\omega ,F)=1\!\!1_{\{\tilde{Z}_\Lambda
^{\sigma ^\tau ,\phi }<\infty \}}(\omega )[\tilde{Z}_\Lambda ^{\sigma ^\tau
,\phi }(\omega )]^{-1}\{1\!\!1_F(\omega _{X\backslash \Lambda }) \\
&+&\sum_{n=1}^\infty \frac{z^n}{n!}\int_{(\Lambda \times
S)^n}\!\!\!\!\!1\!\!1_F(\omega _{X\backslash \Lambda }\cup \{\hat{x}%
\}_1^n)\exp [-\beta E^\phi (\omega _{X\backslash \Lambda }\cup \{\hat{x}%
\}_1^n)]\sigma ^\tau (d\hat{x})_1^n\},
\end{eqnarray*}
where 
\[
\tilde{Z}_\Lambda ^{\sigma ^\tau ,\phi }(\omega )=1+\sum_{n=1}^\infty \frac{%
z^n}{n!}\int_{(\Lambda \times S)^n}\exp [-\beta E^\phi (\omega _{X\backslash
\Lambda }\cup \{\hat{x}\}_1^n)]\sigma ^\tau (d\hat{x})_1^n. 
\]

\item  From properties (S2) and (S3) a probability measure $\mu $ on $%
(\Omega ,\mathcal{B}(\Omega ))$ is a grand canonical Gibbs measure iff for
all $\Lambda \in \mathcal{B}_c(X)$ and all $Y\in \mathcal{B}(\Omega )$ 
\[
\QTR{mathbb}{E}_\mu [{1\!\!1}_Y|\mathcal{B}_{X\backslash \Lambda }(\Omega
)]=\Pi _\Lambda ^{\sigma ^\tau ,\phi }(\cdot ,Y)\;\mu -a.e., 
\]
where for a sub-$\sigma $-algebra $\Sigma \subset \mathcal{B}(\Omega )$, $%
\QTR{mathbb}{E}_\mu [\cdot |\Sigma ]$ denotes the conditional expectation
with respect to $\mu $ given $\Sigma $.
\end{enumerate}
\end{remark}

Furthermore, we want to state the notion of correlation functions.

\begin{definition}
\label{6eq28}For any $m\in \QTR{mathbb}{N}$ and $\Lambda \in \mathcal{B}%
_c(X) $ we define the $m$-point correlation function $\rho _\Lambda
^{(m)}:\Omega _\Lambda ^{(m)}\rightarrow \QTR{mathbb}{R}$ (with empty
boundary condition) by 
\[
\rho _\Lambda ^{(m)}(\{\hat{x}\}_1^m;\emptyset ):=\frac 1{\tilde{Z}_\Lambda
^{z\sigma ^\tau ,\phi }(\emptyset )}\sum_{n=0}^\infty \frac{z^n}{n!}%
\int_{(\Lambda \times S)^n}e^{-\beta E^\phi (\{\hat{x}\}_1^m\cup \{\hat{y}%
\}_1^n)}\sigma ^{\tau \otimes n}(d\hat{y})_1^n. 
\]
\end{definition}

We now formulate the conditions on the interaction which will be used in the
next section.

\begin{enumerate}
\item[(S)]  (Stability) There exists $B\geq 0$ such that 
\begin{equation}
E^\phi (\omega )\geq -B|\omega |,\;\mathrm{for\;any\;}\omega \in \Omega
_{fin}.  \label{6eq114}
\end{equation}

\item[(I)]  (Integrability) We assume the following integrability condition: 
\begin{equation}
C(\beta ):=\stackunder{(y,t)\in X\times S}{\mathrm{ess\,sup}}%
\int_X\int_S|e^{-\beta \phi \left( (x,s),(y,t)\right) }-1|\tau (x,ds)\sigma
(dx)<\infty .  \label{6eq16}
\end{equation}
\end{enumerate}

Only in Theorem~\ref{6eq87} we also need the following notion

\begin{enumerate}
\item[(F)]  (Finite range) There exists $R>0$ such that 
\begin{equation}
\phi ((x,s),(y,t))=0,\quad \mathrm{if\,\,\,}d(x,y)\geq R  \label{6eq132}
\end{equation}
where $d$ denotes the Riemannian distance on $X$.
\end{enumerate}

\begin{remark}
\begin{enumerate}
\item  In the case $X=\QTR{mathbb}{R}^d,$ $S=\{s\}$, $\sigma $ Lebesgue
measure on $\QTR{mathbb}{R}^d$, $\tau (x,\cdot )=\delta _s$ Dirac-measure,
and for translation invariant potential $\phi $ the above integrability
condition (I) reduces to the standard integrability condition, see e.g., 
\cite{R69}.

\item  The stability condition (\ref{6eq114}) implies that for every $\omega
\in \Omega _{fin}$ there is $\hat{x}_0\in \omega $ such that 
\begin{equation}
\sum_{\hat{x}\in \omega \backslash \{\hat{x}_0\}}\phi (\hat{x}_0,\hat{x}%
)\geq -2B,  \label{6eq15}
\end{equation}
in particular, $\phi $ is bounded below.
\end{enumerate}
\end{remark}

\subsection{Examples\label{6eq88}}

Below we give some examples which illustrate different kinds of marked
spe\-cifications arising in models of statistical physics. These examples
can be handled in our framework and we will give more details on the
construction of the marked Gibbs measures corresponding to them in
Subsection~\ref{6eq124}.

\begin{example}
\label{6eq112}Let $X=\QTR{mathbb}{R}^d$, $S=\QTR{mathbb}{R}^l$ be given. As
intensity measure $\sigma $ we take the Lebesgue measure on $\QTR{mathbb}{R}%
^d$ and the kernel $\tau $ is independent of the position and has support in
a compact set.

The potential $\phi $ is given by 
\[
\phi ((x,s_x),(y,s_y)):=\Phi (|x-y|)+J(|x-y|)s_xs_y,
\]
where $\Phi ,J$ are measurable functions on $\Bbb{R}$, such that exist a $%
R,\varepsilon ,C_1,C_2>0$ with $\Phi (r)\geq C_1r^{-d}$ for all $r\leq R$
and $|\Phi |\leq C_2r^{-d-\varepsilon }$ for all $r>R$. $J$ is positive,
decreasing with the distance and for some $a>0$%
\begin{equation}
\sum_{x\in \QTR{mathbb}{Z}^d}J(|ax|)<\infty .  \label{6eq98}
\end{equation}
This model describes a ferromagnetic interaction in fluids, cf.~%
\cite[Sect.~I]{RZ98}. The authors showed the breaking of the discrete
symmetry corresponding to the spin.
\end{example}

\begin{example}
\label{6eq115}Let $X=\QTR{mathbb}{R}^d$ be endowed with Lebesgue measure $dx$
and $S=\QTR{mathbb}{T}$ be the one dimensional torus with measure $\tau $
given by $\tau (x,d\theta _x)=\frac 1{2\pi }d\theta _x$ (we parametrize the
torus by $\theta _x\in [0,2\pi )$). The potential is of the following form 
\[
\phi ((x,\theta _x),(y,\theta _y)):=\Phi (|x-y|)-J(|x-y|)\cos (\theta
_x-\theta _y), 
\]
where $\Phi ,J$ are measurable functions on $\QTR{mathbb}{R}^d$, $J\geq 0$
which fullfil the following conditions: there exist $R,\varepsilon
,C_1,C_2,C_3>0$ such that

\begin{enumerate}
\item  $\Phi \left( r\right) -|J\left( r\right) |\geq C_1r^{-d}$ for all $%
r\leq R$.

\item  $|\Phi \left( r\right) |\leq C_2r^{-d-\varepsilon }$ for all $r>R$.

\item  $|J\left( r\right) |\leq C_3r^{-d-\varepsilon }$ for all $r>R$.
\end{enumerate}

This model describes a classical gas of planar rotators.
\end{example}

\begin{example}
\label{6eq116}We consider as $X$ the Euclidean space $\QTR{mathbb}{R}^d$
with Lebesgue measure and the space of marks $S=\{1,\ldots ,q\}$. The
potential is given by 
\[
\phi ((x,s_x),(y,s_y)):=\varphi (x-y)(1-\delta _{s_x,s_y})+\psi (x-y), 
\]
where $\delta _{s_x,s_y}$ is the Kronecker symbol and $\varphi ,\psi :%
\QTR{mathbb}{R}^d\rightarrow ]-\infty ,\infty ]$ are measurable functions.
We assume that there exist $0\leq r_1\leq r_2$ such that

\begin{enumerate}
\item[(A1)]  (repulsion of $\varphi $) $\varphi \geq 0$.

\item[(A2)]  (finite range of $\varphi $) $\varphi (x)=0$ when $|x|\geq r_2$.

\item[(A3)]  (strong stability and regularity of $\psi $) either $\psi \geq
0 $, or $\psi $ is superstable and lower regular in the sense of \cite{R70}.

\item[(A4)]  the positive part $\psi _{+}$ of $\psi $ satisfies 
\[
\int_{\{x||x|\geq r_1\}}\psi _{+}(x)dx<\infty . 
\]
\end{enumerate}

This model is known as continuum Potts model, cf.~\cite{GH96}.
\end{example}

\begin{example}
\label{6eq126}Let $\mathcal{L}^\theta (\QTR{mathbb}{R}^d)$ be the Banach
space of all continuous functions\newline
$s:[0,\theta ]\rightarrow \QTR{mathbb}{R}^d$ with $s(0)=s(\theta )$ and $%
\theta =\frac 1{k_BT}$, $k_B$ denotes the Boltzmann constant and $T$ the
temperature. On $\mathcal{L}^\theta (\QTR{mathbb}{R}^d)$ we consider the
measure $W_{x,x}(ds)dx$, where $dx$ denotes the Lebesgue measure on $\Bbb{R}%
^d$ and $W_{x,x}(ds)$ the conditional Wiener measure, which is concentrated
on the trajectories starting and ending in $x\in \QTR{mathbb}{R}^d$. In this
framework the potential is of the form 
\[
\phi (s_1,s_2):=\int_0^\theta V(s_1(t)-s_2(t))dt, 
\]
where $V\in L^1(\QTR{mathbb}{R}^d)$ and satisfies 
\[
\sum_{i=1}^n\sum_{j=i+1}^nV(x_i-x_j)\geq -Bn,\;\forall x_1,\ldots x_n\in 
\QTR{mathbb}{R}^d. 
\]
Our aim is to handle the loop space as a marked configuration space putting $%
X=\QTR{mathbb}{R}^d$ equipped with the Lebesgue measure. It would be natural
to consider as mark space at the point $x\in X$ the space $\mathcal{L}%
_x^\theta (X)$ of all loops starting and ending in $x$. In our setting we
are forced to put $S=\mathcal{L}^\theta (X)$. A point in $s\in \mathcal{L}%
^\theta (X)$ is then interpreted as a pair $(s(0),s)$ and denoted by $%
(x,s_x) $. The kernel $\tau $ is given by $\tau (x,ds_x):=W_{x,x}(ds_x)$.
This implies, in particular, that the space $\mathcal{L}_x^\theta (X)$ has
full $\tau \left( x,\cdot \right) $-measure. In Subsection~\ref{6eq124} we
will consider this in more details.

This model is related to the path space representation of the states in
quantum statistical mechanics for Maxwell-Boltzmann statistics. A beautiful
description of this connection for the standard density matrices is given in 
\cite{G71a}. Ginibre also considers the cases of the Bose-Einstein and
Fermi-Dirac statistics. Ginibre does not use any concept of Euclidean Gibbs
measure in his considerations, rather he introduce special version of
correlations functions for which he constructed cluster expansion, etc. The
concept of Euclidean Gibbs measures in quantum statistics was introduced in
the paper \cite{KLRRS97}. This example~shows that it is natural to interpret
such objects as marked Gibbs measures.
\end{example}

\section{Cluster expansion\label{6eq4}}

In this section we derive the cluster expansion of the Gibbs factor $%
e^{-\beta E^{\phi }(\omega )}$, see (\ref{6eq26}) below. Moreover we perform
some estimates which will be used in Section~\ref{6eq5} to prove the
existence of the marked Gibbs measures, cf.~Theorem~\ref{6eq58}.

\subsection{Cluster decomposition property}

\begin{definition}
For any $\omega \in \Omega _{fin}$ we define the function $k$ by 
\begin{equation}
k(\omega ):=\ln ^{*}(e^{-\beta E^\phi })(\omega ),  \label{6eq37}
\end{equation}
or equivalently 
\[
(\exp ^{*}k)(\omega )=e^{-\beta E^\phi (\omega )},
\]
where $E^\phi $ is defined in (\ref{6eq18}). $k$ is called Ursell function
see e.g., \cite{R69}.
\end{definition}

\begin{proposition}
The partition function $\tilde{Z}_\Lambda ^{\sigma ^\tau ,\phi }(\emptyset )$%
, $\Lambda \in \mathcal{B}_c(X)$ has the following representation 
\begin{equation}
\tilde{Z}_\Lambda ^{\sigma ^\tau ,\phi }(\emptyset )=\exp \left(
\int_{\Omega _\Lambda \backslash \{\emptyset \}}k(\omega )\nu _{z\sigma
^\tau }(d\omega )\right) ,  \label{6eq49}
\end{equation}
if $k\in L^1(\Omega _\Lambda ,\nu _{z\sigma ^\tau })$, c.f. Corollary~\ref
{6eq95} below.
\end{proposition}

\noindent \textbf{Proof.} This result follows from the fact that $e^{-\beta
E^\phi (\omega )}=(\exp ^{*}k)(\omega )$ and Corollary~\ref{6eq40}.\hfill $%
\blacksquare $

\begin{proposition}
\label{6eq111}The Ursell function $k$ allows the following representation 
\[
k(\omega )=\sum_{G\in \frak{G}^c(\omega )}\prod_{\{\hat{x},\hat{y}\}\in
G}(e^{-\beta \phi (\hat{x},\hat{y})}-1),\;\omega \in \Omega _{fin}\backslash
\{\emptyset \}, 
\]
and $k(\emptyset )=0$.
\end{proposition}

\noindent \textbf{Proof.} According to the definition of the energy $E^{\phi
}$ (cf.~(\ref{6eq18})) we have 
\begin{eqnarray}
e^{-\beta E^{\phi }(\omega )} &=&\prod_{\{\hat{x},\hat{y}\}\subset \omega
}e^{-\beta \phi (\hat{x},\hat{y})}  \nonumber \\
&=&\sum_{G\in \frak{G}(\omega )}\prod_{\{\hat{x},\hat{y}\}\in G}(e^{-\beta
\phi (\hat{x},\hat{y})}-1).  \label{6eq22}
\end{eqnarray}
Recall that every graph $G$ can be decomposed into the direct sum of its
connected components, i.e., $G=\bigoplus_{i=1}^{n}G_{i}$, where $G_{i}$ is a
connected subgraph and $\{V(G_{i})\}$ is a partition of $\omega $,
(cf.~Subsection \ref{6eq118}). This yields 
\begin{equation}
e^{-\beta E^{\phi }(\omega )}=\sum_{n=0}^{\infty }\frac{1}{n!}\sum_{(\omega
_{1},\ldots ,\omega _{n})\in \frak{P}^{n}(\omega
)}\prod_{l=1}^{n}\sum_{G_{l}\in \frak{G}^{c}(\omega _{l})}\prod_{\{\hat{x},%
\hat{y}\}\in G_{l}}(e^{-\beta \phi (\hat{x},\hat{y})}-1).  \label{6eq24}
\end{equation}

Define $\tilde{k}$ for any $\omega \in \Omega _{fin}\backslash \{\emptyset
\} $ by 
\begin{equation}
\tilde{k}(\omega ):=\sum_{G\in \frak{G}^c(\omega )}\prod_{\{\hat{x},\hat{y}%
\}\in G}(e^{-\beta \phi (\hat{x},\hat{y})}-1),  \label{6eq25}
\end{equation}
and $\tilde{k}(\emptyset )=0$. The expression for $e^{-\beta E^\phi (\omega
)}$ can be written as 
\[
e^{-\beta E^\phi (\omega )}=\sum_{n=0}^\infty \frac 1{n!}\sum_{(\omega
_1,\ldots ,\omega _n)\in \frak{P}^n(\omega )}\tilde{k}(\omega _1)\dots 
\tilde{k}(\omega _n), 
\]
which is nothing but $\exp ^{*}\tilde{k}$. Thus $\tilde{k}=k$ and the result
follows.\hfill $\blacksquare $

\begin{remark}
The equality 
\begin{equation}
e^{-\beta E^\phi (\omega )}=\sum_{n=0}^\infty \frac 1{n!}\sum_{(\omega
_1,\ldots ,\omega _n)\in \frak{P}^n(\omega )}k(\omega _1)\dots k(\omega _n),
\label{6eq26}
\end{equation}
is known as the cluster decomposition of the Gibbs factor $e^{-\beta E^\phi
(\omega )}$. We also notice that the function $k$ is $\mathcal{B}(\Omega
_{fin})$-measurable.
\end{remark}

Proposition~\ref{6eq111} allows the following representation for the
specification 
\[
\Pi _\Lambda ^{\sigma ^\tau ,\phi }(\emptyset ,F)=\frac 1{\tilde{Z}_\Lambda
^{\sigma ^\tau ,\phi }(\emptyset )}\int_{\Omega _\Lambda }1\!\!1_F(\omega
)\exp ^{*}(k)(\omega )\nu _{z\sigma ^\tau }(d\omega ),\quad F\in \mathcal{B}%
(\Omega ) 
\]
where $\tilde{Z}_\Lambda ^{\sigma ^\tau ,\phi }(\emptyset )$ is given by (%
\ref{6eq49}).

The next proposition gives a relation between correlations functions (see
Definition~\ref{6eq28}) and Ursell functions.

\begin{proposition}
\label{6eq123}Let $\omega \in \Omega _\Lambda ^{(m)}$, $\Lambda \in \mathcal{%
B}_c(X)$ be given. Define 
\[
\bar{k}(\omega ,\omega ^{\prime }):=(\exp ^{*}(-k)*D_\omega e^{-\beta E^\phi
})(\omega ^{\prime }), 
\]
for $\gamma _\omega \cap \gamma _{\omega ^{\prime }}\neq \emptyset $. If $%
k\in L^1(\Omega _\Lambda ,\nu _{z\sigma ^\tau })$ (cf.~(\ref{6eq63}) below),
then 
\[
\rho _\Lambda ^{(m)}(\omega ;\emptyset )=\int_{\Omega _\Lambda }\bar{k}%
(\omega ,\omega ^{\prime })\nu _{z\sigma ^\tau }(d\omega ^{\prime }). 
\]
\end{proposition}

\noindent \textbf{Proof.} It follows from the definition of $\rho _\Lambda
^{(m)}$ and (\ref{6eq49}) that 
\[
\rho _\Lambda ^{(m)}(\omega ;\emptyset )=\exp \left( -\int_{\Omega _\Lambda
}k(\omega ^{\prime })\nu _{z\sigma ^\tau }(d\omega ^{\prime })\right)
\int_{\Omega _\Lambda }(D_\omega e^{-\beta E^\phi })(\omega ^{\prime })\nu
_{z\sigma ^\tau }(d\omega ^{\prime }). 
\]
Now taking into account Lemma \ref{6eq38} with $F=1$ and Corollary \ref
{6eq40} the above equality gives 
\[
\int_{\Omega _\Lambda }(\exp ^{*}(-k)*D_\omega e^{-\beta E^\phi })(\omega
^{\prime })\nu _{z\sigma ^\tau }(d\omega ^{\prime }). 
\]
\hfill $\blacksquare $

Let us now derive an explicit relation between $k$ and $\bar{k}$.

\begin{lemma}
Let $\omega \in \Omega _{fin}\backslash \{\emptyset \}$ be given and suppose
that $\hat{x}\in \omega $. Then $k$ and $\bar{k}$ are related by the
equation 
\begin{equation}
\bar{k}(\{\hat{x}\},\omega \backslash \{\hat{x}\})=k(\omega ).  \label{6eq51}
\end{equation}
\end{lemma}

\noindent \textbf{Proof.} By definition of $\bar{k}$ and Proposition~\ref
{6eqT1}-3 we have 
\begin{eqnarray*}
\bar{k}(\{\hat{x}\},\omega \backslash \{\hat{x}\}) &=&(\exp ^{*}(-k)*D_{\{%
\hat{x}\}}\exp ^{*}(k))(\omega \backslash \{\hat{x}\}) \\
&=&(\exp ^{*}(-k)*\exp ^{*}(k)*D_{\{\hat{x}\}}k)(\omega \backslash \{\hat{x}%
\}) \\
&=&(D_{\{\hat{x}\}}k)(\omega \backslash \{\hat{x}\}) \\
&=&k(\omega ).
\end{eqnarray*}
Hence the result is proved.\hfill $\blacksquare $

\begin{remark}
Notice that for any $\omega ,\omega ^{\prime }\in \Omega _{fin}\backslash
\{\emptyset \}$ with $\gamma _\omega \cap \gamma _{\omega ^{\prime
}}=\emptyset $, $\bar{k}$ may be written as 
\[
\bar{k}(\omega ,\omega ^{\prime })=\sum_{l=1}^\infty \frac
1{l!}\sum_{(\omega _1,\ldots ,\omega _l)\in \frak{P}^l(\omega
)}\sum_{(\omega _1^{\prime },\ldots ,\omega _l^{\prime })\in \frak{P}%
_\emptyset ^l(\omega ^{\prime })}k(\omega _1\cup \omega _1^{\prime })\ldots
k(\omega _l\cup \omega _l^{\prime }), 
\]
which is the same as the sum of all graphs where each connected component
has at least one vertex in the points of $\omega $, cf.~\cite[Chap.~4]{MM91}.
\end{remark}

Our aim now is to find a bound for $\bar{k}$. First we derive an equation of
recursive type for $\bar{k}$. Let $\omega ,\zeta \in \Omega _{fin}$ be such
that $\gamma _\omega \cap \gamma _\zeta =\emptyset $ and $\hat{x}_0$ an
arbitrary fixed element in $\omega $. To this end we look again into the
definition of $\bar{k}$ which can be expressed as 
\begin{equation}
\bar{k}(\omega ,\zeta )=\sum_{\omega ^{\prime }\subset \zeta }(\exp
^{*}(-k))(\zeta \backslash \omega ^{\prime })D_{\{\hat{x}_0\}}e^{-\beta
E^\phi (\omega \backslash \{\hat{x}_0\}\cup \omega ^{\prime })}.
\label{6eq52}
\end{equation}
Having in mind the decomposition (\ref{6eq77}) for $E^\phi $ one obtains
(taking into account (\ref{6eq78})) 
\begin{eqnarray*}
e^{-\beta W(\{\hat{x}_0\},\omega ^{\prime })} &=&\prod_{\hat{x}\in \omega
^{\prime }}e^{-\beta \phi (\hat{x}_0,\hat{x})} \\
&=&\sum_{\omega ^{\prime \prime }\subset \omega ^{\prime }}\prod_{\hat{x}\in
\omega ^{\prime \prime }}(e^{-\beta \phi (\hat{x}_0,\hat{x})}-1) \\
&=&\sum_{\omega ^{\prime \prime }\subset \omega ^{\prime }}k_{\omega
^{\prime \prime }}(\hat{x}_0),
\end{eqnarray*}
where 
\[
k_{\omega ^{\prime \prime }}(\hat{x}_0):=\prod_{\hat{x}\in \omega ^{\prime
\prime }}(e^{-\beta \phi (\hat{x}_0,\hat{x})}-1). 
\]
According to equation (\ref{6eq52}) $\bar{k}$ can be formulated as 
\begin{eqnarray*}
\bar{k}(\omega ,\zeta ) &=&e^{-\beta W(\{\hat{x}_0\},\omega \backslash \{%
\hat{x}_0\})}\sum_{\omega ^{\prime }\subset \zeta }(\exp ^{*}(-k))(\zeta
\backslash \omega ^{\prime }) \\
&&\times \sum_{\omega ^{\prime \prime }\subset \omega ^{\prime }}k_{\omega
^{\prime \prime }}(\hat{x}_0)e^{-\beta E^\phi (\omega \backslash \{\hat{x}%
_0\}\cup \omega ^{\prime })}.
\end{eqnarray*}
Interchanging the two sums the right hand side becomes 
\[
e^{-\beta W(\{\hat{x}_0\},\omega \backslash \{\hat{x}_0\})}\sum_{\omega
^{\prime \prime }\subset \zeta }k_{\omega ^{\prime \prime }}(\hat{x}_0)\sum_{%
\QATOP{\omega ^{\prime }}{\omega ^{\prime \prime }\subset \omega ^{\prime
}\subset \xi }}(\exp ^{*}(-k))(\zeta \backslash \omega ^{\prime })e^{-\beta
E^\phi (\omega \backslash \{\hat{x}_0\}\cup \omega ^{\prime })}, 
\]
and the second sum may be rewritten as 
\begin{eqnarray*}
&&\sum_{\check{\omega}\subset \zeta \backslash \omega ^{\prime \prime
}}(\exp ^{*}(-k))(\zeta \backslash (\check{\omega}\cup \omega ^{\prime
\prime }))e^{-\beta E^\phi (\omega \backslash \{\hat{x}_0\}\cup \check{\omega%
}\cup \omega ^{\prime \prime })} \\
&=&\sum_{\check{\omega}\subset \zeta \backslash \omega ^{\prime \prime
}}(\exp ^{*}(-k))((\zeta \backslash \omega ^{\prime \prime })\backslash 
\check{\omega})(D_{\omega \backslash \{\hat{x}_0\}\cup \omega ^{\prime
\prime }}e^{-\beta E^\phi })(\check{\omega}).
\end{eqnarray*}
Therefore $\bar{k}$ can be expressed as 
\begin{eqnarray*}
\bar{k}(\omega ,\zeta ) &=&e^{-\beta W(\{\hat{x}_0\},\omega \backslash \{%
\hat{x}_0\})}\sum_{\omega ^{\prime \prime }\subset \zeta }k_{\omega ^{\prime
\prime }}(\hat{x}_0) \\
&&\times (\exp ^{*}(-k)*(D_{\omega \backslash \{\hat{x}_0\}\cup \omega
^{\prime \prime }}e^{-\beta E^\phi })(\zeta \backslash \omega ^{\prime
\prime }).
\end{eqnarray*}
Finally, taking into account the definition of $\bar{k}$ we arrive at 
\begin{equation}
\bar{k}(\omega ,\zeta )=e^{-\beta W(\{\hat{x}_0\},\omega \backslash \{\hat{x}%
_0\})}\sum_{\omega ^{\prime }\subset \zeta }k_{\omega ^{\prime }}(\hat{x}_0)%
\bar{k}(\omega \backslash \{\hat{x}_0\}\cup \omega ^{\prime },\zeta
\backslash \omega ^{\prime }).  \label{6eq53}
\end{equation}
According to the definition of $\bar{k}$ we have for the case $\omega
=\emptyset $ that $\bar{k}(\emptyset ,\zeta )=1^{*}(\zeta ),$ where $1^{*}$
is defined by (\ref{6eq79}).

\subsection{Convergence of cluster expansion}

We want to derive now a bound for $|\bar{k}(\omega ,\zeta )|$ which will be
used later on in the main estimation in this section (cf.~\ref{6eq140}). The
idea is to define a function $Q$ dominating $\bar{k}$ which fulfills an
equation similar to (\ref{6eq53}) which can be solved, see Proposition~\ref
{6eq12} below.

Let us choose a mapping $I:\Omega _{fin}\rightarrow X\times S,$ $\tilde{%
\omega}\mapsto I(\tilde{\omega})\in \tilde{\omega}$ such that the following
equation is fulfilled 
\begin{equation}
\sum_{\hat{x}\in \tilde{\omega}\backslash I(\tilde{\omega})}\phi (\hat{x},I(%
\tilde{\omega}))>-2B.  \label{6eq80}
\end{equation}
Such a mapping exists by the stability condition, see (\ref{6eq15}).

Of course given $I$ and $\bar{k}$, (\ref{6eq53}) implies 
\begin{equation}
\bar{k}(\omega ,\zeta )=\exp \left( -\beta \sum_{\hat{x}\in \omega
\backslash I(\omega )}\phi (\hat{x},I(\omega ))\right) \sum_{\omega ^{\prime
}\subset \zeta }k_{\omega ^{\prime }}(I(\omega ))\bar{k}(\omega \backslash
I(\omega )\cup \omega ^{\prime },\zeta \backslash \omega ^{\prime }).
\label{6eq121}
\end{equation}

Now we can start defining $Q_I$ inductively. For $\omega =\emptyset $ we
define 
\begin{equation}
Q(\emptyset ,\zeta ):=1^{*}(\zeta ),  \label{6eq55}
\end{equation}
and by definition of $\bar{k}(\emptyset ,\zeta )$ we have $|\bar{k}%
(\emptyset ,\zeta )|\leq Q_I(\emptyset ,\zeta )$.

Assume we already have defined $Q_I$ for all $\omega ,\zeta \in \Omega
_{fin} $, $\omega \neq \emptyset $, $\gamma _\omega \cap \gamma _\zeta
=\emptyset $, and $|\omega \cup \zeta |=n$ such that 
\[
|\bar{k}(\omega ,\zeta )|\leq Q_I(\omega ,\zeta ) 
\]
is satisfied. Then if $\omega ,\zeta \in \Omega _{fin}$, are such that $%
\omega \neq \emptyset $, $\gamma _\omega \cap \gamma _\zeta =\emptyset $,
and $|\omega \cup \zeta |=n+1$, we have, applying (\ref{6eq80}) and (\ref
{6eq121}) 
\[
|\bar{k}(\omega ,\zeta )|\leq e^{2\beta B}\sum_{\omega ^{\prime }\subset
\zeta }|k_{\omega ^{\prime }}(I(\omega ))|Q_I(\omega \backslash I(\omega
)\cup \omega ^{\prime },\zeta \backslash \omega ^{\prime }). 
\]
Thus we define 
\begin{equation}
Q_I(\omega ,\zeta ):=e^{2\beta B}\sum_{\omega ^{\prime }\subset \zeta
}|k_{\omega ^{\prime }}(I(\omega ))|Q_I(\omega \backslash I(\omega )\cup
\omega ^{\prime },\zeta \backslash \omega ^{\prime }),  \label{6eq54}
\end{equation}

\begin{remark}
The solutions of the equations (\ref{6eq53}) and (\ref{6eq54}) exist and are
unique. Let us explain this in more details. On the one hand the equations
are linear, on the other hand the value at the point $(\omega ,\zeta )$ for $%
|\omega |+|\zeta |=n$ only depends on the values at points $(\tilde{\omega},%
\tilde{\zeta})$ with $|\tilde{\omega}|+|\tilde{\zeta}|=n-1$, thus the
corresponding matrix is an strict upper triangle matrix for a suitable
choice of the bases.
\end{remark}

Hence we have the following proposition.

\begin{proposition}
For $I$ and $\bar{k}$ as above, there exists a unique solution $Q_I$ of the
equation (\ref{6eq54}) with the initial condition (\ref{6eq55}) which
dominates $\bar{k}$, i.e., for any $\omega ,\zeta \in \Omega _{fin}$, such
that $\gamma _\omega \cap \gamma _\zeta =\emptyset $ we have $|\bar{k}%
(\omega ,\zeta )|\leq Q_I(\omega ,\zeta )$.
\end{proposition}

The next proposition gives a solution for the equation (\ref{6eq54}) which
does not depend on the choice of $I$.

\begin{proposition}
\label{6eq12}Let $\omega ,\zeta \in \Omega _{fin}$ with $\gamma _\omega \cap
\gamma _\zeta =\emptyset $. The solution of (\ref{6eq54}) for $\omega =\{%
\hat{x}_1,\ldots ,\hat{x}_l\}$, $l\geq 1$ has the form 
\begin{equation}
Q(\{\hat{x}_1,\ldots ,\hat{x}_l\},\zeta )=\sum_{(\omega _1,\ldots ,\omega
_l)\in \frak{P}_\emptyset ^l(\zeta )}Q(\{\hat{x}_1\},\omega _1)\cdots Q(\{%
\hat{x}_l\},\omega _l),  \label{6eq84}
\end{equation}
where 
\begin{equation}
Q(\{\hat{x}\},\zeta ):=(e^{2\beta B})^{|\zeta |+1}\sum_{T\in \frak{T}(\{\hat{%
x}\}\cup \zeta )}\prod_{\{\hat{y},\hat{y}^{\prime }\}\in T}|e^{-\beta \phi (%
\hat{y},\hat{y}^{\prime })}-1|,  \label{6eq10}
\end{equation}
for $\zeta \neq \emptyset $ and $Q(\{\hat{x}\},\emptyset ):=e^{2\beta B}$.
In the case $\omega =\emptyset $ we define $Q(\emptyset ,\zeta )$ as in (\ref
{6eq55}).
\end{proposition}

The proof of this proposition is notationally quite involved because of the
``reordering'' of graphs, therefore we give the details in Appendix~\ref
{6eq72}.

As a result we have the following proposition.

\begin{proposition}
\label{6eq129}For any $\omega ,\zeta \in \Omega _{fin}$ such that $\omega
\neq \emptyset $, $\gamma _\omega \cap \gamma _\zeta =\emptyset $, and $%
\omega =\{\hat{x}_1,\ldots ,\hat{x}_l\}$, $l\geq 1$ we have 
\begin{eqnarray*}
|\bar{k}(\omega ,\zeta )| &\leq &Q(\{\hat{x}_1,\ldots ,\hat{x}_l\},\zeta ) \\
&=&\sum_{(\zeta _1,\ldots ,\zeta _l)\in \frak{P}_\emptyset ^l(\zeta )}Q(\{%
\hat{x}_1\},\zeta _1)\cdots Q(\{\hat{x}_l\},\zeta _l),
\end{eqnarray*}
and 
\begin{equation}
|k(\omega )|\leq e^{2\beta B|\omega |}\sum_{T\in \frak{T}(\omega )}\prod_{\{%
\hat{x},\hat{x}^{\prime }\}\in T}|e^{-\beta \phi (\hat{x},\hat{x}^{\prime
})}-1|.  \label{6eq56}
\end{equation}
\end{proposition}

\noindent \textbf{Proof.} The first part follows from the previous
proposition and the second part follows from the relation between $k$ and $%
\bar{k}$ (cf.~(\ref{6eq51})) and (\ref{6eq10}).\hfill $\blacksquare $

Using the fact that the sum in the function $Q$ is only over trees we can
give also estimates for integrals.

\begin{lemma}
\label{6eq128}For every $\hat{x}\in X\times S$, $Y\in \mathcal{B}(X)$, and $%
n\geq 1$ we have 
\begin{eqnarray}
&&\int_{(Y\times S)^n}Q(\{\hat{x}\},\{\hat{y}\}_1^n)\sigma ^\tau (d\hat{y}%
)_1^n  \label{6eq140} \\
&\leq &e^{2\beta B(n+1)}C(\beta )^{n-1}(n+1)^{n-1}\int_{Y\times S}|e^{-\beta
\phi (\hat{x},\hat{y})}-1|\sigma ^\tau (d\hat{y}).  \nonumber
\end{eqnarray}
\end{lemma}

\noindent We refer to the Appendix~\ref{6eq75} for the proof of this lemma.

\begin{proposition}
\label{6eq74}Let $\Lambda \in \mathcal{B}_c(\Omega )$ be given. Then for any 
$z$ such that 
\[
|z|<\frac 1{2e}(e^{2\beta B}C(\beta ))^{-1}, 
\]
where $C(\beta )$ is given by integrability condition (\ref{6eq16}), we have 
\begin{equation}
\int_{\Omega _\Lambda \backslash \{\emptyset \}}\int_{\Omega
_{fin}}|k(\omega \cup \omega ^{\prime })|\nu _{z\sigma ^\tau }(d\omega )\nu
_{z\sigma ^\tau }(d\omega ^{\prime })<\infty .  \label{6eq50}
\end{equation}
\end{proposition}

\noindent \textbf{Proof: }Using the definition of $\nu _{z\sigma ^\tau }$
and the relation between $k$ and $\overline{k}$ (cf. (\ref{6eq51})) we may
write (\ref{6eq50}) as 
\[
\sum_{n=1}^\infty \sum_{m=0}^\infty \frac{z^{n+m}}{n!m!}\int_{(\Lambda
\times S)^n}\int_{(X\times S)^m}|\overline{k}(\{x_n\},\{\hat{x}%
\}_1^{n-1}\cup \{\hat{y}\}_1^m)|\sigma ^\tau (d\hat{x})_1^n\sigma ^\tau (d%
\hat{y})_1^m.
\]
According to Proposition~\ref{6eq129} and Lemma \ref{6eq128} we can bound
the above term by 
\begin{equation}
\leq \sum_{n=1}^\infty \sum_{m=0}^\infty \frac{z^{n+m}}{n!m!}e^{2\beta
B(n+m)}C(\beta )^{n+m-2}(n+m)^{n+m-2}C(\beta )\int_{(\Lambda \times
S)}\sigma ^\tau (d\hat{x}_n).  \label{6eq32}
\end{equation}
Using the fact that $(n+m)^{n+m-2}\leq e^{m+n}(m+n)!$ (c.f. Remark~\ref
{6eq131}.1.) we estimate (\ref{6eq32}) by 
\begin{eqnarray*}
&&\frac{\int_\Lambda \tau (x,S)\sigma (dx)}{C(\beta )}\sum_{m=1}^\infty
\sum_{n=0}^\infty \frac{(m+n)!}{m!n!}(zeC(\beta )e^{2\beta B})^{m+n} \\
&=&\frac{\int_\Lambda \tau (x,S)\sigma (dx)}{C(\beta )}\sum_{l=0}^\infty
\left( \sum_{m=1}^l\frac{l!}{m!(l-m)!}\right) (zeC(\beta )e^{2\beta B})^l \\
&\leq &\frac{\int_\Lambda \tau (x,S)\sigma (dx)}{C(\beta )}\sum_{l=0}^\infty
(2zeC(\beta )e^{2\beta B})^l,
\end{eqnarray*}
from which the result follows.\hfill $\blacksquare $

As a consequence of the last proposition and Fubini's theorem we have the
following corollary.

\begin{corollary}
\label{6eq95}For any $\Lambda \in \mathcal{B}_c(X)$ we have (notice that $%
k(\emptyset )=0$, see Proposition~\ref{6eq111}) 
\begin{equation}
\int_{\Omega _\Lambda }|k(\omega )|\nu _{z\sigma ^\tau }(d\omega )<\infty ,
\label{6eq63}
\end{equation}
and for $\nu _{z\sigma ^\tau }$-a.a.~$\omega ^{\prime }\in \Omega
_{fin}\backslash \{\emptyset \}$%
\begin{equation}
\int_{\Omega _{fin}}|k(\omega \cup \omega ^{\prime })|\nu _{z\sigma ^\tau
}(d\omega )<\infty .  \label{6eq64}
\end{equation}
\end{corollary}

\section{Construction of marked Gibbs measure \label{6eq5}}

\subsection{Limiting measures from cluster expansion\label{6eq122}}

Below we construct the marked measure $\mu $ on $\Omega $ as a limiting
measure of the specification $\Pi _\Lambda ^{\sigma ^\tau ,\phi }$ for the
empty boundary condition in the weak local sense (cf.~Theorem~\ref{6eq58}).
Some sets of full $\mu $-measure are considered (cf. Proposition~\ref{6eq127}%
). For the case that the potential $\phi $ has finite range we also give an
easy proof that the resulting limiting measure satisfies the DLR equations,
cf.~Theorem~\ref{6eq87}. All results extend to separable standard Borel
spaces. This is explained in some details in Subsection~\ref{6eq76}.
Finally, we show in Subsection~\ref{6eq124} how to apply the abstract
results to the examples given in Subsection~\ref{6eq88}.

\begin{lemma}
\label{6eq81}For any $\Lambda ,\Lambda ^{\prime }\in \mathcal{B}_c(X)$ such
that $\Lambda ^{\prime }\subset \Lambda $ and $F\in \mathcal{B}(\Omega
_{\Lambda ^{\prime }})$ the specification $\Pi _\Lambda ^{\sigma ^\tau ,\phi
}(\emptyset ,p_{\Lambda ^{\prime }}^{-1}(F))$ has the following
representation 
\begin{eqnarray*}
&&\Pi _\Lambda ^{\Lambda ^{\prime }}(\emptyset ,F):=\Pi _\Lambda ^{\sigma
^\tau ,\phi }(\emptyset ,p_{\Lambda ^{\prime }}^{-1}(F)) \\
&=&\frac 1{\tilde{Z}_\Lambda ^{\Lambda ^{\prime }}(\emptyset )}\int_{\Omega
_{\Lambda ^{\prime }}}\!\!\!\!1\!\!1_F(\omega )\exp ^{*}\left(
1\!\!1_{\Omega _{fin}\backslash \{\emptyset \}}(\cdot )\int_{\Omega
_{\Lambda \backslash \Lambda ^{\prime }}}\!\!\!\!\!\!\!k(\omega ^{\prime
}\cup \cdot )\nu _{z\sigma ^\tau }(d\omega ^{\prime })\right) \!(\omega )\nu
_{z\sigma ^\tau }(d\omega ),
\end{eqnarray*}
where 
\[
\tilde{Z}_\Lambda ^{\Lambda ^{\prime }}(\emptyset ):=\exp \left(
\int_{\Omega _{\Lambda ^{\prime }}\backslash \{\emptyset \}}\int_{\Omega
_{\Lambda \backslash \Lambda ^{\prime }}}\!\!\!k(\omega \cup \omega ^{\prime
})\nu _{z\sigma ^\tau }(d\omega ^{\prime })\nu _{z\sigma ^\tau }(d\omega
)\right) .
\]
\end{lemma}

\noindent \textbf{Proof.} This is a result of the following direct
calculation 
\begin{eqnarray*}
\Pi _\Lambda ^{\sigma ^\tau ,\phi }(\emptyset ,p_{\Lambda ^{\prime
}}^{-1}(F)) &=&[\tilde{Z}_\Lambda ^{\sigma ^\tau ,\phi }(\emptyset
)]^{-1}\int_{\Omega _\Lambda }1\!\!1_{p_{\Lambda ^{\prime }}^{-1}(F)}(\omega
^{\prime })\exp [-\beta E^\phi (\omega ^{\prime })]\nu _{z\sigma ^\tau
}(d\omega ^{\prime }) \\
&=&[\tilde{Z}_\Lambda ^{\sigma ^\tau ,\phi }(\emptyset )]^{-1}\int_{\Omega
_{\Lambda ^{\prime }}}1\!\!1_F(\omega ) \\
&&\times \int_{\Omega _{\Lambda \backslash \Lambda ^{\prime }}}\exp [-\beta
E^\phi (\omega \cup \omega ^{\prime })]\nu _{z\sigma ^\tau }(d\omega
^{\prime })\nu _{z\sigma ^\tau }(d\omega ).
\end{eqnarray*}
Because $k\in L^1(\Omega _\Lambda ,\nu _{z\sigma ^\tau })$ ( cf. Corollary~%
\ref{6eq95}), using Lemma \ref{6eq41} the inner integral can be rewritten as
follows 
\begin{eqnarray*}
&&\int_{\Omega _{\Lambda \backslash \Lambda ^{\prime }}}\exp [-\beta E^\phi
(\omega \cup \omega ^{\prime })]\nu _{z\sigma ^\tau }(d\omega ^{\prime
})=\int_{\Omega _{\Lambda \backslash \Lambda ^{\prime }}}\left( \exp
^{*}k\right) (\omega \cup \omega ^{\prime })\nu _{z\sigma ^\tau }(d\omega
^{\prime }) \\
&=&\exp \left( \int_{\Omega _{\Lambda \backslash \Lambda ^{\prime
}}}\!\!\!\!\!\!\!k(\omega ^{\prime })\nu _{z\sigma ^\tau }(d\omega ^{\prime
})\right) \exp ^{*}\left( 1\!\!1_{\Omega _{fin}\backslash \{\emptyset
\}}(\cdot )\int_{\Omega _{\Lambda \backslash \Lambda ^{\prime
}}}\!\!\!\!\!\!k(\omega ^{\prime }\cup \cdot )\nu _{z\sigma ^\tau }(d\omega
^{\prime })\right) \!\!(\omega ).
\end{eqnarray*}
\hfill $\blacksquare $

\begin{proposition}
\label{6eq59}Let $\Lambda ,\Lambda ^{\prime }\in \mathcal{B}_c(X)$ be such
that $\Lambda ^{\prime }\subset \Lambda $.

\begin{enumerate}
\item  \label{6eq60}Let $k_\Lambda ^{\Lambda ^{\prime }}$ be defined by 
\begin{equation}
k_\Lambda ^{\Lambda ^{\prime }}(\omega ):=1\!\!1_{\Omega _{fin}\backslash
\{\emptyset \}}(\omega )\int_{\Omega _{\Lambda \backslash \Lambda ^{\prime
}}}\!k(\omega \cup \omega ^{\prime })\nu _{z\sigma ^\tau }(d\omega ^{\prime
}).  \label{6eq67}
\end{equation}
Then for $\nu _{z\sigma ^\tau }$-a.a.$~\omega \in \Omega _{fin}\backslash
\{\emptyset \}$ we have $\lim_{\Lambda \nearrow X}k_\Lambda ^{\Lambda
^{\prime }}(\omega )=k^{\Lambda ^{\prime }}(\omega )$, where 
\begin{equation}
k^{\Lambda ^{\prime }}(\omega )=1\!\!1_{\Omega _{fin}\backslash \{\emptyset
\}}(\omega )\int_{\Omega _{X\backslash \Lambda ^{\prime
},fin}}\!\!\!\!\!k(\omega \cup \omega ^{\prime })\nu _{z\sigma ^\tau
}(d\omega ^{\prime }).  \label{6eq61}
\end{equation}

\item  \label{6eq62}We have also that $\lim_{\Lambda \nearrow X}\tilde{Z}%
_\Lambda ^{\Lambda ^{\prime }}(\emptyset )=\tilde{Z}^{\Lambda ^{\prime
}}(\emptyset )$, where 
\begin{equation}
\tilde{Z}^{\Lambda ^{\prime }}(\emptyset )=\exp \left( \int_{\Omega
_{\Lambda ^{\prime }}\backslash \{\emptyset \}}\!\!\!k^{\Lambda ^{\prime
}}(\omega )\nu _{z\sigma ^\tau }(d\omega )\right) >0.  \label{6eq66}
\end{equation}
\end{enumerate}
\end{proposition}

\noindent \textbf{Proof.} \ref{6eq60}. We would like to estimate the
following quantity 
\[
\left| k_\Lambda ^{\Lambda ^{\prime }}(\omega )-\int_{\Omega _{X\backslash
\Lambda ^{\prime },fin}}k(\omega \cup \omega ^{\prime })\nu _{z\sigma ^\tau
}(d\omega ^{\prime })\right| . 
\]
According to the definition of $k_\Lambda ^{\Lambda ^{\prime }}$ the above
quantity is estimated by 
\begin{equation}
\int_{\Omega _{X\backslash \Lambda ^{\prime },fin}\backslash \Omega
_{\Lambda \backslash \Lambda ^{\prime }}}|k(\omega \cup \omega ^{\prime
})|\nu _{z\sigma ^\tau }(d\omega ^{\prime }).  \label{6eq65}
\end{equation}
Now let $\{\Lambda _n|n\in \QTR{mathbb}{N\}}$ be a sequence of increasing
volumes such that $\Lambda _n\nearrow X$. Then $\Omega _{\Lambda
_n\backslash \Lambda ^{\prime }}\nearrow \Omega _{X\backslash \Lambda
^{\prime },fin}$. On the other hand (\ref{6eq64}) guarantees that there
exists a $\nu _{z\sigma ^\tau }$-null set $N\in \mathcal{B}(\Omega _{fin})$
such that for all $\omega \in \Omega _{fin}\backslash (N\cup \{\emptyset \})$%
\[
\int_{\Omega _{fin}}|k(\omega \cup \omega ^{\prime })|\nu _{z\sigma ^\tau
}(d\omega ^{\prime })<\infty . 
\]
Therefore by Lebesgue's dominated convergence theorem it follows that (\ref
{6eq65}) goes to zero for all fixed $\omega \in \Omega _{fin}\backslash
(N\cup \{\emptyset \})$. Hence part~\ref{6eq60} is proved.

To prove part~\ref{6eq62} we note that 
\[
|k_\Lambda ^{\Lambda ^{\prime }}(\omega )|\leq \int_{\Omega
_{fin}}\!\!\!|k(\omega \cup \omega ^{\prime })|\nu _{z\sigma ^\tau }(d\omega
^{\prime }), 
\]
and thus 
\begin{equation}
\int_{\Omega _{\Lambda ^{\prime }}\backslash \{\emptyset
\}}\!\!\!\!\!|k_\Lambda ^{\Lambda ^{\prime }}(\omega )|\nu _{z\sigma ^\tau
}(d\omega )\leq \int_{\Omega _{\Lambda ^{\prime }}\backslash \{\emptyset
\}}\int_{\Omega _{fin}}\!\!\!|k(\omega \cup \omega ^{\prime })|\nu _{z\sigma
^\tau }(d\omega ^{\prime })\nu _{z\sigma ^\tau }(d\omega )<\infty ,
\label{6eq82}
\end{equation}
because of (\ref{6eq50}). This implies that 
\[
\lim_{\Lambda \nearrow X}\int_{\Omega _{\Lambda ^{\prime }}\backslash
\{\emptyset \}}k_\Lambda ^{\Lambda ^{\prime }}(\omega )\nu _{z\sigma ^\tau
}(d\omega )=\int_{\Omega _{\Lambda ^{\prime }}\backslash \{\emptyset
\}}\int_{\Omega _{X\backslash \Lambda ^{\prime },fin}}\!\!\!\!k(\omega \cup
\omega ^{\prime })\nu _{z\sigma ^\tau }(d\omega ^{\prime })\nu _{z\sigma
^\tau }(d\omega ), 
\]
and, of course, taking exponential, and having in mind the form of $%
k^{\Lambda ^{\prime }}$ in (\ref{6eq61}), the desired result (\ref{6eq66})
follows.\hfill $\blacksquare $

\begin{theorem}
\label{6eq58}For any $z$ such that $|z|<\frac 1{2e}(e^{2\beta B}C(\beta
))^{-1}$, where $C(\beta )$ is given by integrability condition (\ref{6eq16}%
), the specification $\Pi _\Lambda ^{\sigma ^\tau ,\phi }(\emptyset ,d\omega
)$ converges in the weak local \allowbreak sense to a measure $\mu $, i.e.,
for any bounded $\mathcal{B}_{\Lambda ^{\prime }}(\Omega )$-measurable
function $F$ we have 
\[
\int_\Omega F(\omega )\Pi _\Lambda ^{\sigma ^\tau ,\phi }(\emptyset ,d\omega
)\rightarrow \frac 1{\tilde{Z}^{\Lambda ^{\prime }}(\emptyset )}\int_{\Omega
_{\Lambda ^{\prime }}}F(\omega )(\exp ^{*}k^{\Lambda ^{\prime }})(\omega
)\nu _{z\sigma ^\tau }(d\omega ), 
\]
and thus 
\begin{equation}
\mu ^{\Lambda ^{\prime }}(d\omega )=\frac 1{\tilde{Z}^{\Lambda ^{\prime
}}(\emptyset )}(\exp ^{*}k^{\Lambda ^{\prime }})(\omega )\nu _{z\sigma ^\tau
}(d\omega ).  \label{6eq104}
\end{equation}
\end{theorem}

\noindent \textbf{Proof.} Let $F$ be as stated above, then Lemma \ref{6eq81}
implies 
\[
\int_\Omega \!\!F(\omega )\Pi _\Lambda ^{\sigma ^\tau ,\phi }(\emptyset
,d\omega )=\frac 1{\tilde{Z}_\Lambda ^{\Lambda ^{\prime }}(\emptyset
)}\!\int_{\Omega _{\Lambda ^{\prime }}}\!\!\!\!\!f(\omega
)\!\sum_{n=0}^\infty \frac 1{n!}\!\sum_{(\omega _1,\ldots ,\omega _n)\in 
\frak{P}^n(\omega )}\prod_{i=1}^nk_\Lambda ^{\Lambda ^{\prime }}(\omega
_i)\nu _{z\sigma ^\tau }(d\omega ), 
\]
where $F=f\circ p_{\Lambda ^{\prime }}$. According to Proposition~\ref{6eq59}
we know already that $\tilde{Z}_\Lambda ^{\Lambda ^{\prime }}(\emptyset )$
converges to $\tilde{Z}^{\Lambda ^{\prime }}(\emptyset )$. In order to use
the Lebesgue dominated convergence theorem one should estimate the integrand
by a function which is integrable and independent of $\Lambda .$ An
appropriate bound is 
\begin{eqnarray*}
&&\sum_{n=0}^\infty \frac 1{n!}\sum_{(\omega _1,\ldots ,\omega _n)\in \frak{P%
}^n(\omega )}\prod_{i=1}^n\int_{\Omega _{X\backslash \Lambda ^{\prime
},fin}}\!\!\!|k(\zeta \cup \omega _i)|\nu _{z\sigma ^\tau }(d\zeta ) \\
&=&\exp ^{*}\left( {1\!\!1}_{\Omega _{fin}\backslash \{\emptyset \}}(\cdot
)\int_{\Omega _{X\backslash \Lambda ^{\prime },fin}}\!\!\!\!\!|k(\zeta \cup
\cdot )|\nu _{z\sigma ^\tau }(d\zeta )\right) (\omega ).
\end{eqnarray*}
Moreover, the integral of the bound is given by 
\begin{eqnarray*}
&&\int_{\Omega _{\Lambda ^{\prime }}}\exp ^{*}\left( {1\!\!1}_{\Omega
_{fin}\backslash \{\emptyset \}}(\cdot )\int_{\Omega _{X\backslash \Lambda
^{\prime },fin}}\!\!\!\!\!|k(\zeta \cup \cdot )|\nu _{z\sigma ^\tau }(d\zeta
)\right) (\omega )\nu _{z\sigma ^\tau }(d\omega ) \\
&=&\exp \left( \int_{\Omega _{\Lambda ^{\prime }}\backslash \{\emptyset
\}}\int_{\Omega _{X\backslash \Lambda ^{\prime },fin}}\!\!\!\!\!|k(\zeta
\cup \omega )|\nu _{z\sigma ^\tau }(d\zeta )\nu _{z\sigma ^\tau }(d\omega
)\right) <\infty ,
\end{eqnarray*}
because of Corollary \ref{6eq40} and (\ref{6eq50}). Therefore we have the
desired result 
\[
\lim_{\Lambda \nearrow X}\int_{\Omega _{\Lambda ^{\prime }}}F(\omega )\Pi
_\Lambda ^{\sigma ^\tau ,\phi }(\emptyset ,d\omega )=\int_{\Omega _{\Lambda
^{\prime }}}F(\omega )\frac 1{Z^{\Lambda ^{\prime }}(\emptyset )}(\exp
^{*}k^{\Lambda ^{\prime }})(\omega )\nu _{z\sigma ^\tau }(d\omega ). 
\]
\hfill $\blacksquare $

The measure from Theorem~\ref{6eq58} is not concentrated in all $\Omega $,
indeed we have the following:

\begin{proposition}
\label{6eq127}Let $A$ be a $\mathcal{B}(X\times S)$-measurable set such that 
$\sigma ^\tau (A)=0$, then the set of marked configurations not touching $A$%
, i.e., 
\[
\tilde{\Omega}=\{\omega =(\gamma ,s)\in \Omega |(x,s_x)\in A^c,\forall x\in
\gamma \}, 
\]
has full $\mu $-measure.
\end{proposition}

\noindent \textbf{Proof:} Let us prove that $\mu (\tilde{\Omega}^c)=0$. To
this end we write $\tilde{\Omega}^c$ as 
\begin{eqnarray*}
\tilde{\Omega}^c &=&\{\omega =(\gamma ,s)\in \Omega |(x,s_x)\in A,\;\mathrm{%
for\;some\;}x\in \gamma \} \\
&=&\bigcup_{n\in \QTR{mathbb}{N}}p_{\Lambda _n}^{-1}(\{\omega =(\gamma
,s)\in \Omega _{\Lambda _n}|(x,s_x)\in A,\;\mathrm{for\;some\;}x\in \gamma
_{\Lambda _n}\}).
\end{eqnarray*}
Therefore 
\[
\mu (\tilde{\Omega}^c)\leq \sum_{n=1}^\infty \mu ^{\Lambda _n}(\{\omega
=(\gamma ,s)\in \Omega _{\Lambda _n}|(x,s_x)\in A,\;\mathrm{for\;some\;}x\in
\gamma _{\Lambda _n}\}). 
\]
Since $\mu ^{\Lambda _n}\ll \nu _{z\sigma ^\tau }$ (cf.(\ref{6eq104})), it
is enough to prove that 
\[
\nu _{z\sigma ^\tau }(\{\omega =(\gamma ,s)\in \Omega |(x,s_x)\in A,\;%
\mathrm{for\;some\;}x\in \gamma _{\Lambda _n}\})=0. 
\]
According to the definition of $\nu _{z\sigma ^\tau }$ (cf.~(\ref{6eq93}))
the left hand side of the above equality yields 
\begin{eqnarray*}
&&\sum_{m=0}^\infty \frac{z^m}{m!}\sigma _m^\tau (\{(\hat{x}_1,\ldots ,\hat{x%
}_m)\in (\Lambda _n\times S)^m/S_m|\hat{x}_i\in A\;\mathrm{for\;some\;}i\})
\\
&\leq &\sum_{m=0}^\infty \frac{z^mm}{m!}(\sigma ^\tau (\Lambda _n\times
S))^{m-1}\sigma ^\tau (A),
\end{eqnarray*}
which is zero since $\sigma ^\tau (A)=0$.\hfill $\blacksquare $

\begin{remark}
Since the projections of tempered Gibbs measures at arbitrary temperature
and fugacity are absolutely continuous with respect to the Lebesgue Poisson
measure (cf.~\cite{R69} and \cite{KoKu98}) the above considerations extends
also to all Gibbs measures.
\end{remark}

\subsection{Identification with Gibbs measures\label{6eq130}}

If we additionally assume finite range of the potential, then it is a direct
consequence that the limit measure from Theorem$~$\ref{6eq58} verifies the
DLR equations. Without this additional assumption a more detailed
consideration is necessary, see. \cite{KoKu98} and \cite{Ku98}.

\begin{theorem}
\label{6eq87}For any finite range potential $\phi $ (c.f. \ref{6eq132}) the
measure $\mu $ from Theorem~\ref{6eq58} fulfils the DLR equations.
\end{theorem}

\noindent \textbf{Proof.} Let $\Lambda \in \mathcal{B}_c(X)$ be given. Then
there exists a $\tilde{\Lambda}$ $\in \mathcal{B}_c(X)$ such that $\Lambda
\subset \tilde{\Lambda}$ and $\phi ((x,s_x),(y,s_y))=0$ if $x\in \Lambda $
and $y\in \tilde{\Lambda}^c$. Whence the interaction energy is 
\[
W(\omega _\Lambda ,\omega _{X\backslash \Lambda })=\sum_{\hat{x}\in \omega
_\Lambda }\sum_{\hat{y}\in \omega _{X\backslash \Lambda }}\phi (\hat{x},\hat{%
y})=\sum_{\hat{x}\in \omega _\Lambda }\sum_{\hat{y}\in \omega _{\tilde{%
\Lambda}\backslash \Lambda }}\phi (\hat{x},\hat{y})=W(\omega _\Lambda
,\omega _{\tilde{\Lambda}\backslash \Lambda }), 
\]
and the sums are finite. Let $F$ be a ``locally'' measurable set, i.e.,
there exists $\widehat{\Lambda }\in \mathcal{B}_c(X)$ with $F\in \mathcal{B}%
_{\widehat{\Lambda }}(\Omega )$, then 
\[
\Pi _\Lambda ^{\sigma ^\tau ,\phi }(\omega ,F)=\frac{1\!\!1_{\{\tilde{Z}%
_\Lambda ^{\sigma ^\tau ,\phi }<\infty \}}(\omega _{\tilde{\Lambda}%
\backslash \Lambda })}{\tilde{Z}_\Lambda ^{\sigma ^\tau ,\phi }(\omega _{%
\tilde{\Lambda}\backslash \Lambda })}\int_{\Omega _\Lambda }1\!\!1_F(\zeta
\cup \omega _{X\backslash \Lambda })e^{-\beta E_\Lambda ^\phi (\zeta \cup
\omega _{\tilde{\Lambda}\backslash \Lambda })}\nu _{z\sigma ^\tau }(d\zeta
), 
\]
which implies that $\Pi _\Lambda ^{\sigma ^\tau ,\phi }(\cdot ,F)$ is a
bounded $\mathcal{B}_{\tilde{\Lambda}\cup \widehat{\Lambda }}(\Omega )$%
-measurable function. Additionally, we have for all $\Lambda ^{\prime }\in 
\mathcal{B}_c(X)$ with $\Lambda \subset \Lambda ^{\prime }$%
\[
\int_\Omega \Pi _\Lambda ^{\sigma ^\tau ,\phi }(\omega ,F)\Pi _{\Lambda
^{\prime }}^{\sigma ^\tau ,\phi }(\emptyset ,d\omega )=\Pi _{\Lambda
^{\prime }}^{\sigma ^\tau ,\phi }(\emptyset ,F), 
\]
(cf.~Remark~\ref{6eq117}-(S4)). Since $\Pi _{\Lambda ^{\prime }}^{\sigma
^\tau ,\phi }(\emptyset ,\cdot )\rightarrow \mu $ in the weak local sense
also 
\[
\int_\Omega \Pi _\Lambda ^{\sigma ^\tau ,\phi }(\omega ,F)\Pi _{\Lambda
^{\prime }}^{\sigma ^\tau ,\phi }(\emptyset ,d\omega )\rightarrow
\int_\Omega \Pi _\Lambda ^{\sigma ^\tau ,\phi }(\omega ,F)\mu (d\omega ) 
\]
for $\Lambda ^{\prime }\nearrow X$. Moreover, $\Pi _{\Lambda ^{\prime
}}^{\sigma ^\tau ,\phi }(\emptyset ,F)\rightarrow \mu (F)$ which implies the
DLR equations 
\[
\int_\Omega \Pi _\Lambda ^{\sigma ^\tau ,\phi }(\omega ,F)\mu (d\omega )=\mu
(F). 
\]
\hfill $\blacksquare $

\subsection{Extension to standard Borel spaces\label{6eq76}}

In this subsection we will present a natural generalization of our results.
Except in Theorem~\ref{6eq58} and \ref{6eq87} we use nothing else than the
measurability structure of $X$ and $S$ and there we only apply the theorem
of Kolmogorov for projective limits. Thus the construction works as well for 
$X$ and $S$ separable standard Borel spaces. To this end we recall the
definition and properties of separable standard Borel space, see e.g., \cite
{C83}, \cite{G88} and \cite{P67}.

\begin{definition}
Let $\left( X,\frak{F}\right) $ and $\left( X^{\prime },\frak{F}^{\prime
}\right) $ be two measurable spaces.

\begin{enumerate}
\item  The spaces $\left( X,\frak{F}\right) $ and $\left( X^{\prime },\frak{F%
}^{\prime }\right) $ are called isomorphic iff there exists a measurable
bijective mapping $f:X\rightarrow X^{\prime }$ such that its inverse $f^{-1}$
is also measurable.

\item  $\left( X,\frak{F}\right) $ and $\left( X^{\prime },\frak{F}^{\prime
}\right) $ are called $\sigma $-isomorphic iff there exists a bijective
mapping $F:\frak{F}\rightarrow \frak{F}^{\prime }$ between the $\sigma $%
-algebras which preserves the operations in a $\sigma $-algebra.

\item  $\left( X,\frak{F}\right) $ is said to be countable generated iff
there exists a denumerable class $\frak{D}\subset \frak{F}$ such that $\frak{%
D}$ generates $\frak{F}$.

\item  $\left( X,\frak{F}\right) $ is said to be separable iff it is
countably generated and for each $x\in X$ the set $\left\{ x\right\} \in 
\frak{F}$.
\end{enumerate}
\end{definition}

\begin{definition}
\label{6eq97}Let $\left( X,\frak{F}\right) $ be a countably generated
measurable space. Then $\left( X,\frak{F}\right) $ is called standard Borel
space iff there exists a Polish space $(X^{\prime },\frak{F}^{\prime })$
(i.e., metrizable, complete metric space which fulfills the second axiom of
countability and the $\sigma $-algebra $\frak{F}^{\prime }$ coincides with
the Borel $\sigma $-algebra) such that $(X,\frak{F})$ and $(X^{\prime },%
\mathcal{B}\left( X^{\prime }\right) )$ are $\sigma $-isomorphic.
\end{definition}

\begin{example}
\begin{enumerate}
\item  Every locally compact, $\sigma $-compact space is a standard Borel
space.

\item  Polish spaces are standard Borel spaces.
\end{enumerate}
\end{example}

We have the following proposition, cf.~\cite[Chap.~V, Theorem~2.1]{P67}.

\begin{proposition}
\begin{enumerate}
\item  If $\left( X,\frak{F}\right) $ is a countable generated measurable
space, then there exists $E\subset \left\{ 0,1\right\} ^{\QTR{mathbb}{N}}$
such that $(X,\frak{F})$ is $\sigma $-isomorphic to $(E,\mathcal{B}(E))$.
Thus $\left( X,\frak{F}\right) $ is $\sigma $-isomorphic to a separable
measurable space.

\item  Let $\left( X,\frak{F}\right) $ and $\left( X^{\prime },\frak{F}%
^{\prime }\right) $ be separable measurable spaces. Then $\left( X,\frak{F}%
\right) $ is $\sigma $-isomorphic to $\left( X^{\prime },\frak{F}^{\prime
}\right) $ iff they are isomorphic.
\end{enumerate}
\end{proposition}

Finally we state some operations under which separable standard Borel space
are closed, see e.g., \cite{P67} and \cite{C83}.

\begin{theorem}
\label{6eq101}Let $(X_1,\frak{F}_1),(X_2,\frak{F}_2),\ldots $ be separable
standard Borel spaces.

\begin{enumerate}
\item  \label{6eq103}Countable product, sums, and union are separable
standard Borel spa\-ces.

\item  The projective limit is a separable standard Borel space.

\item  Any measurable subset of a separable standard Borel space is also a
separable standard Borel space.
\end{enumerate}
\end{theorem}

We need also a version of Kolmogorov's theorem for separable standard Borel
spaces.

\begin{theorem}
\label{6eq100}(cf.~\cite[Chap.~V Theorem~3.2]{P67}) Let $(X_n,\frak{F}_n)$, $%
n\in \Bbb{N}$ be separable standard Borel spaces. Let $(X,\frak{F})$ be the
projective limit of the space $(X_n,\frak{F}_n)$ relative to the maps $%
p_{n,m}:X_n\rightarrow X_m$, $m\leq n$. If $\{\mu _n\}_{n\in \Bbb{N}}$ is a
sequence of probability measures such that $\mu _n$ is a measure on $(X_n,%
\frak{F}_n)$ and $\mu _m=\mu _n\circ p_{n,m}^{-1}$ for $m\leq n$; then there
exists a unique measure $\mu $ on $(X,\frak{F})$ such that $\mu _n=\mu \circ
p_n^{-1}$ for all $n\in \Bbb{N}$ where $p_n$ is the projection map from $X$
on $X_n$.
\end{theorem}

This theorem can be extended to an index set $I$ which is a directed set
with an order generating sequence, i.e., there exists a sequence $(\alpha
_n)_{n\in \Bbb{N}}$ in $I$ such that for every $\alpha \in I$ exists a $n\in 
\Bbb{N}$ with $\alpha <\alpha _n$.

Let us now apply this general framework to our marked configuration space $%
\Omega $.

We assume, therefore, that $(X,\frak{X})$, $(S,\frak{S})$ are separable
standard Borel spaces.

To use $\mathcal{B}_{c}(X)$ makes in this generality no sense, hence we have
to introduce an abstract concept of ``local'' sets. Let $\frak{I}_{X}$ be a
subset of $\frak{X}$ with the properties:

\begin{enumerate}
\item[(I1)]  $\Lambda _1\cup \Lambda _2\in \frak{I}_X$ for all $\Lambda
_1,\Lambda _2\in \frak{I}_X$.

\item[(I2)]  If $\Lambda \in \frak{I}_X$ and $A\in \frak{X}$ with $A\subset
\Lambda $ then $A\in \frak{I}_X$.

\item[(I3)]  There exists a sequence $\{\Lambda _n,n\in \QTR{mathbb}{N}\}$
from $\frak{I}_X$ with $X=\bigcup_{n\in \QTR{mathbb}{N}}\Lambda _n$ and such
that if $\Lambda \in \frak{I}_X$ then $\Lambda \subset \Lambda _n$ for some $%
n\in \QTR{mathbb}{N}$.
\end{enumerate}

\noindent Then we can construct the marked configuration space as in
Subsection~\ref{6eq6} replacing $\mathcal{B}_{c}(X)$ by $\frak{I}_{X}$. Our
aim is to show that $(\Omega ,\mathcal{B}(\Omega )$ is a separable standard
Borel space and thus by Theorem~\ref{6eq100} the measure in Theorem~\ref
{6eq58} exists.

It follows from Theorem~\ref{6eq101} that for any $\Lambda \in \frak{I}_X$
and any $n\in \QTR{mathbb}{N}$ the set $(\Lambda \times S)^n$ is a separable
standard Borel space. Therefore, by the same argument $(\widetilde{\Lambda
\times S})^n/S_n$ is also a separable standard Borel space, see e.g., \cite
{S94}. Now taking into account the isomorphism (cf.~\ref{6eq102}) between $(%
\widetilde{\Lambda \times S})^n/S_n$ and $\Omega _\Lambda ^{(n)}$ the same
holds for $\Omega _\Lambda ^{(n)}$, Hence $\Omega _\Lambda $ is also a
separable standard Borel space as well as $\Omega _X^{(n)}$ by Theorem~\ref
{6eq101}, (\ref{6eq103}).

Finally, the marked configuration space itself is a separable standard Borel
space as the projective limit of the separable standard Borel spaces $%
(\Omega _{\Lambda },\mathcal{B}(\Omega _{\Lambda }))$, $\Lambda \in \frak{I}%
_{X}$.

Furthermore, if on $(X,\frak{X})$ is given a non-atomic measure $\sigma $
with $\sigma (\Lambda )<\infty $ $\forall \Lambda \in \frak{I}_X$ and a
kernel $\tau :X\times \frak{G}\rightarrow \QTR{mathbb}{R}$ which fulfills (%
\ref{6eq57}), then the procedure from Subsection~\ref{6eq92} can be done in
an analogous way and as a result we obtain a probability measure $\pi
_{z\sigma }^\tau $, $z>0$ on $(\Omega ,\mathcal{B}(\Omega ))$.
Specifications and marked Gibbs measures can also be defined analogously,
see e.g., \cite{P80}. All the contents of Sections \ref{6eq4} and \ref{6eq5}
generalize straightforward and only in Theorem~\ref{6eq58} we need the
assumption that the spaces are standard Borel. In Theorem~\ref{6eq87} we
have to generalize the notion of finite range, actually we use in the proof
only the following property: For any $\Lambda \in \frak{I}_X$ exists a $%
\Lambda ^{\prime }\in \frak{I}_X$ such that $\Lambda \subset \Lambda
^{\prime }$ and $\phi ((x,s),(y,t))=0$ if $x\in \Lambda $ and $y\in
X\backslash \Lambda ^{\prime }$.

\subsection{Examples revisited\label{6eq124}}

Here we will verify that our framework is sufficient to treat the examples
stated in Subsection~\ref{6eq88}. Therefore the main task in this subsection
is to verify the stability condition (S) (cf.~(\ref{6eq114})) and
integrability condition (I) (cf.~(\ref{6eq16})) for each of the examples in
Subsection~\ref{6eq88}. This enables us to apply Theorem~\ref{6eq58} to the
considered examples and therefore give us a limit measures corresponding to
the specifications under consideration.

\begin{proposition}
Let $\phi $ be a potential on $X\times S$ bounded below, then the following
three conditions are equivalent

\begin{enumerate}
\item[(i)]  the potential fulfills the integrability condition.

\item[(ii)]  there exists a $\alpha >0$ such that for 
\[
A_\alpha \mbox{$:=$}\{(x,s_x)\in X\times S|\,\phi ((y,s_y),(x,s_x))>\alpha \}
\]
the following is fulfilled: 
\[
\stackunder{(y,s_y)\in X\times S}{\mathrm{ess\,sup}}\sigma ^\tau (A_\alpha
)<+\infty 
\]
and 
\[
\stackunder{(y,s_y)\in X\times S}{\mathrm{ess\,sup}}\int_{(X\times
S)\backslash A_\alpha }\!\!\!\!\!\!|\phi ((y,s_y),(x,s_x))|\,\tau
(x,ds_x)\sigma (dx)<+\infty .
\]

\item[(iii)]  For every $\hat{y}\in X\times S$ there exists a $N_{\hat{y}%
}\in \mathcal{B}(X)\otimes \mathcal{B}(S)$ such that 
\[
\stackunder{\hat{y}\in X\times S}{\mathrm{ess\,sup}}\sigma ^\tau (N_{\hat{y}%
})<+\infty ,
\]
and 
\[
\stackunder{(y,s_y)\in X\times S}{\mathrm{ess\,sup}}\int_{\left( X\times
S\right) \backslash N_{\hat{y}}}\!\!|\phi ((y,s_y),(x,s_x))|\,\tau
(x,ds_x)\sigma (dx)<+\infty .
\]
\end{enumerate}

In particular, integrability condition (I) (cf.~(\ref{6eq16})) is
independent of $\beta $.
\end{proposition}

\noindent \textbf{Proof.} Denote the lower bound for $\phi $ by $B^{\prime }.
$ Using the fact that there exist constants $C_1,C_2>0$ such that 
\begin{eqnarray*}
C_1\left( |x|1\!\!1_{[-B^{\prime },\alpha )}(x)+1\!\!1_{(\alpha ,\infty
)}(x)\right)  &\leq &1\!\!1_{[-B^{\prime },\infty )}(x)|e^{-x}-1| \\
&\leq &C_2\left( |x|1\!\!1_{[-B^{\prime },\alpha )}(x)+1\!\!1_{(\alpha
,\infty )}(x)\right) ,
\end{eqnarray*}
we see that (i) and (ii) are equivalent. Obviously, (ii) implies (iii) if we
put $N_{\hat{y}}:=\{\hat{x}\in X\times S|\,\phi (\hat{y},\hat{x})\leq \alpha
\}$. Conversely, using 
\begin{eqnarray*}
\{\hat{x} &\in &X\times S|\,\phi (\hat{y},\hat{x})\leq \alpha \} \\
\!\!\!\!\!\!\!\! &=&\left( \{\hat{x}\in X\times S|\,\phi (\hat{y},\hat{x}%
)\leq \alpha \}\cap N_{\hat{y}}\right) \sqcup \left( \{\hat{x}\in X\times
S|\,\phi (\hat{y},\hat{x})\leq \alpha \}\backslash N_{\hat{y}}\right) ,
\end{eqnarray*}
we obtain that also (iii) implies (ii).\hfill $\blacksquare $

Thus we have the following sufficient condition for the integrability
condition.

\begin{corollary}
Let $X$ be a Riemannian manifold ($d$ denotes the metric on $X$). Let us
assume that there exists $R>0$ such that 
\[
\stackunder{(y,s_y)\in X\times S}{\mathrm{ess\,sup}}\int_{\{(x,s_x)\in
X\times S|\,d((y,s_y),(x,s_x))\geq R\}}\!\!\!\!\!\!|\phi
((y,s_y),(x,s_x))|\,\tau (x,ds_x)\sigma (dx)<+\infty .
\]
Then the above conditions are fulfilled.
\end{corollary}

\begin{proposition}
\begin{enumerate}
\item  Let $\phi _1,\phi _2$ be two stable potentials then also $\phi
_1+\phi _2$ is stable.

\item  Let $\phi _1,\phi _2$ be two potentials which fulfil the
integrability condition then $\phi _1+\phi _2$ also satisfies the
integrability condition.

\item  A potential bounded from below by a stable one is stable
itself.\bigskip 
\end{enumerate}
\end{proposition}

\noindent \textbf{Example \ref{6eq112} }$\Phi $ is integrable on $\{|x|\geq
R\}$ and because of monotonicity of $J$ there exists a $C>0$ such that 
\[
\int J(|x|)dx\leq C\sum_{q\in \QTR{mathbb}{Z}^d}J\left( a|q|\right) <+\infty
.
\]
We can bound the potential $\phi $ below by 
\[
\phi ((x,s_x),(y,s_y))\geq \Phi (|x-y|)-K|J(|x-y|)|,
\]
for $K:=\sup_{s\in \limfunc{supp}\tau }|s|^2$ and this potential is stable
according to the Dobrushin-Fisher-Ruelle criterium (cf.~Section~3.2.8 in 
\cite{R69}). This potential fulfills also the integrability condition
because we can bound it above by 
\[
|\phi ((x,s_x),(y,s_y))|\leq |\Phi (|x-y|)|+K|J(|x-y|)|,
\]
and this is integrable on $\{|x|\geq R\}$.\bigskip 

\noindent \textbf{Example \ref{6eq115}} The arguments are analogous to the
above case.

\noindent \textbf{Example \ref{6eq116}} On the one hand in this model the
potential is bounded below by 
\[
\phi ((x,s_x),(y,s_y))\geq \psi (|x-y|), 
\]
and thus stable. On the other hand its modulus is bounded above by 
\[
|\phi ((x,s_x),(y,s_y))|\leq \varphi (|x-y|)+|\psi (|x-y|)|, 
\]
whence it fulfills the integrability condition, because the lower regularity
of $\psi $ implies that also $\psi _{-}$ is integrable.\bigskip

\noindent \textbf{Example \ref{6eq126}} The potential is stable since for
all $\{(x_1,s_1),\ldots ,(x_n,s_n)\}$%
\[
\sum_{i=1}^n\sum_{j=i+1}^n\int_0^\theta V(s_i(t)-s_j(t))dt\geq
-nB\int_0^\theta dt=-nB\theta . 
\]
The potential fulfills the integrability condition because of the following
arguments. Let $s_x\in \mathcal{L}^\theta (\QTR{mathbb}{R}^d)$, $s_y\in 
\mathcal{L}^\theta (\QTR{mathbb}{R}^d)$ and denote by $\tilde{s}%
_x:=s_x-s_x(0)$ and $\tilde{s}_y:=s_y-s_y(0)$ then 
\begin{eqnarray*}
&&\int_{\QTR{mathbb}{R}^d}\int_{\mathcal{L}^\theta (\QTR{mathbb}{R}^d)}|\phi
((x,s_x),(y,s_y))|\tau (x,ds_x)\sigma (dx) \\
&\leq &\int_{\QTR{mathbb}{R}^d}\int_{\mathcal{L}^\theta (\QTR{mathbb}{R}%
^d)}\int_0^\theta |V(x+\tilde{s}_x(t)-s_y(t))|dtW_{0|0}(d\tilde{s}_x)dx \\
&=&\int_{\mathcal{L}^\theta (\QTR{mathbb}{R}^d)}\int_0^\theta \int_{%
\QTR{mathbb}{R}^d}|V(x+\tilde{s}_x(t)-s_y(t))|dxdtW_{0|0}(d\tilde{s}_x) \\
&=&\int_{\QTR{mathbb}{R}^d}|V\left( x\right) |dx\int_{\mathcal{L}^\theta (%
\QTR{mathbb}{R}^d)}\int_0^\theta dtW_{0|0}(d\tilde{s}_x)\leq \frac 1{(2\pi
\theta )^{d/2}}\theta \int_{\QTR{mathbb}{R}^d}|V\left( x\right) |dx.
\end{eqnarray*}
Thus according to Theorem~\ref{6eq58} there exists a limiting measure $\mu $
on the marked configuration space $\Omega _{\QTR{mathbb}{R}^d}(\mathcal{L}%
^\theta (\QTR{mathbb}{R}^d))$, this are not configurations in loops but we
can embed the loop space into $\QTR{mathbb}{R}^d\Bbb{\times }\mathcal{L}%
^\theta (\QTR{mathbb}{R}^d)$ in the following way 
\[
I:\mathcal{L}^\theta (\QTR{mathbb}{R}^d)\hookrightarrow \QTR{mathbb}{R}^d%
\Bbb{\times }\mathcal{L}^\theta (\QTR{mathbb}{R}^d),\;s\mapsto (s(0),s), 
\]
and the image of this mapping 
\[
A:=\{(x,s)\in \QTR{mathbb}{R}^d\Bbb{\times }\mathcal{L}^\theta (\QTR{mathbb}{%
R}^d)|\,s(0)=x\}, 
\]
is a measurable set of full measure, i.e., 
\[
\int_{A^c}W_{x|x}(ds)dx=0. 
\]
Whence according to Proposition~\ref{6eq127} the set 
\[
\tilde{\Omega}:=\{\{(x_1,s_{x_1}),(x_2,s_{x_2}),\ldots \}\in \Omega
|\,\,s_{x_i}(0)=x_i\;\mathrm{for\;\,all}\;\,i\}, 
\]
has full measure, i.e., $\mu (\tilde{\Omega})=1$ and thus we can define a
measure on the loop space via the above embedding.\bigskip

\noindent \textbf{Acknowledgments}\medskip

We would like to thank S.~Albeverio, ~R.~A.~Minlos, and G.~V.~\allowbreak
Shchepan'uk for many helpful discussions. We also thank V.~A.~Zagrebnov for
pointing out Example \ref{6eq112}. Financial support of the INTAS-Project
378, PRAXIS Programme through CITMA, Funchal, Research Center BiBoS, and TMR
Nr.~ERB4001GT957046 is gratefully acknowledged.

\addcontentsline{toc}{section}{Appendix}

\begin{appendix}

\section*{Appendix}

\setcounter{section}{1}

\subsection{Proof of Lemma \ref{6eq41} \label{6eq73}}

\begin{lemma}
\label{6eq120}The following results are valid.

\begin{enumerate}
\item  \label{6eq105}$\sigma ^{\tau \otimes n}(\{(\hat{x}_1,\ldots ,\hat{x}%
_n)\in (X\times S)^n|\exists i,j\;i\neq j\;\mathrm{with}\;x_i=x_j\})=0$.

\item  \label{6eq106}$\sigma ^{\tau \otimes n}((X\times S)^n\backslash (%
\widetilde{X\times S})^n)=0$.

\item  \label{6eqT2}For all $\omega \in \Omega _{fin}\ $the set $A_\omega
:=\{\omega ^{\prime }\in \Omega _{fin}|\gamma _\omega \cap \gamma _{\omega
^{\prime }}\neq \emptyset \}$ has zero $\nu _{z\sigma ^\tau }$-measure.

\item  \label{6eq107}The set $A:=\{(\omega ,\omega ^{\prime })\in \Omega
_{fin}\times \Omega _{fin}|\gamma _\omega \cap \gamma _{\omega ^{\prime
}}\neq \emptyset \}$ has zero $\nu _{z\sigma ^\tau }\otimes \nu _{z\sigma
^\tau }$-measure.
\end{enumerate}
\end{lemma}

\noindent \textbf{Proof.} \ref{6eq105}. Because of the symmetry and the
non-atomicity of $\sigma $ we have 
\begin{eqnarray*}
&&\sigma ^{\tau \otimes n}(\{(\hat{x}_{1},\ldots ,\hat{x}_{n})\in (X\times
S)^{n}|\exists i,j\;i\neq j\;\mathrm{with}\;x_{i}=x_{j}\}) \\
&\leq &\binom{n}{2}\sigma ^{\tau \otimes n}(\{(\hat{x}_{1},\ldots ,\hat{x}%
_{n})\in (X\times S)^{n}|x_{1}=x_{2}\}) \\
&=&\binom{n}{2}\sigma ^{\tau }(X\times S)^{n-2}\sigma ^{\tau \otimes
2}(\{((x,s),(x,t))|x\in X,\,s,t\in S\}) \\
&=&0
\end{eqnarray*}

\ref{6eq106}. Consequence of \ref{6eq105}.

\ref{6eqT2}. Let $\omega =\{y_1,\ldots ,y_m\}$. According to (\ref{6eq89})
we can decompose the set $A_\omega $ as 
\[
A_\omega =\bigsqcup_{n=0}^\infty \{\omega ^{\prime }\in \Omega ^{(n)}|\gamma
_\omega \cap \gamma _{\omega ^{\prime }}\neq \emptyset \}, 
\]
then the definition of $\nu _{z\sigma ^\tau }$ applied to $A$ yields 
\[
\nu _{z\sigma ^\tau }(A)=\sum_{n=0}^\infty \frac{z^n}{n!}(\sigma _n^\tau
)(A_{\omega ,n}), 
\]
where $A_{\omega ,n}$ is given by 
\[
A_{\omega ,n}:=\{\{\hat{x}_1,\ldots ,\hat{x}_n\}\in \Omega
^{(n)}\;|\;\exists i,j\;i\neq j\;\mathrm{with}\;x_i=y_j\}. 
\]
On the other hand we can estimate $\frac 1{n!}(\sigma _n^\tau )(A_{\omega
,n})$ by 
\[
nm\sigma ^{\tau \otimes n}(\{(\hat{x}_1,\ldots ,\hat{x}_n)\in (\widetilde{%
X\times S})^n\times (\widetilde{X\times S})^m|x_1=y_1\}), 
\]
then the definition of $\sigma ^\tau $ and the non-atomicity of $\sigma $
implies as above that this last expression is zero.

\ref{6eq107}. Consequence of \ref{6eqT2}.$\hfill \blacksquare $

\noindent \textbf{Lemma \ref{6eq41}} Let $\psi \in \mathcal{A}$ and $\Lambda
,\Lambda ^{\prime }\in \mathcal{B}_{c}(X)$ be given such that $\Lambda
^{\prime }\subset \Lambda $, suppose that $\psi \in L^{1}(\Omega _{\Lambda
},\nu _{z\sigma ^{\tau }})$. Then the following equality holds 
\begin{eqnarray}
&&\int_{\Omega _{\Lambda \backslash \Lambda ^{\prime }}}(\exp ^{*}\psi
)(\omega \cup \omega ^{\prime })\nu _{z\sigma ^{\tau }}(d\omega )
\label{6eq83} \\
&=&\exp \left( \int_{\Omega _{\Lambda \backslash \Lambda ^{\prime
}}}\!\!\!\!\!\!\psi (\omega )\nu _{z\sigma ^{\tau }}(d\omega )\right) \exp
^{*}\left( \int_{\Omega _{\Lambda \backslash \Lambda ^{\prime
}}}\!\!\!\!\!\!1\!\!1_{\Omega _{fin}\backslash \{\emptyset \}}(\cdot )\psi
(\cdot \cup \omega )\nu _{z\sigma ^{\tau }}(d\omega )\right) (\omega
^{\prime }),  \nonumber
\end{eqnarray}
for $\nu _{z\sigma ^{\tau }}$-a.e.~$\omega \in \Omega _{\Lambda ^{\prime }}$.

\noindent \textbf{Proof.} First we clarify the existence of the integrals.
It follows from Fubini's theorem that 
\begin{eqnarray*}
&&\int_{\Omega _{\Lambda ^{\prime }}}\int_{\Omega _{\Lambda \backslash
\Lambda ^{\prime }}}|(\exp ^{*}\psi )(\omega \cup \omega ^{\prime })|\nu
_{z\sigma ^\tau }(d\omega ^{\prime })\nu _{z\sigma ^\tau }(d\omega ) \\
&\leq &\int_{\Omega _{\Lambda ^{\prime }}}\int_{\Omega _{\Lambda \backslash
\Lambda ^{\prime }}}(\exp ^{*}|\psi |)(\omega \cup \omega ^{\prime })\nu
_{z\sigma ^\tau }(d\omega ^{\prime })\nu _{z\sigma ^\tau }(d\omega ) \\
&=&\exp \left( \int_{\Omega _\Lambda }|\psi |(\omega )\nu _{z\sigma ^\tau
}(d\omega )\right) <\infty ,
\end{eqnarray*}
thus for $\nu _{z\sigma ^\tau }$-a.e.~$\omega \in \Omega _{\Lambda ^{\prime
}}$ $(\exp ^{*}|\psi |)(\cdot \cup \omega ^{\prime })$ and $(\exp ^{*}\psi
)(\cdot \cup \omega ^{\prime })$ belongs to $L^1(\Omega _{\Lambda \backslash
\Lambda ^{\prime }},\nu _{z\sigma ^\tau })$. Hence the following
manipulations are justified for $|\psi |$ and therefore also for the
function $\psi $ itself.

The left hand side of (\ref{6eq83}) for $\omega ^{\prime }\neq \emptyset $
is by definition equivalent to 
\begin{equation}
\int_{\Omega _{\Lambda \backslash \Lambda ^{\prime }}}\sum_{n=1}^{\infty }%
\frac{1}{n!}\sum_{(\omega _{1},\ldots ,\omega _{n})\in \frak{P}_{\emptyset
}^{n}(\omega \cup \omega ^{\prime })}\psi (\omega _{1})\dots \psi (\omega
_{n})\nu _{z\sigma ^{\tau }}(d\omega ).  \label{6eq68}
\end{equation}

\noindent Without loss of generality we may assume $\gamma _\omega \cap
\gamma _{\omega ^{\prime }}=\emptyset $ (cf.~Lemma~\ref{6eq120}). To each
partition $(\omega _1,\ldots ,\omega _n)\in \frak{P}_\emptyset ^n(\omega
\cup \omega ^{\prime })$ we define in one to one form the following objects
(we put together the $\omega _i$'s which have solely points from $\omega $) 
\begin{eqnarray*}
J &\mbox{\rm{$:=$}}&\{i\,|\,\omega _i\subset \omega \} \\
l &\mbox{\rm{$:=$}}&|J| \\
\eta _i &\mbox{\rm{$:=$}}&\omega _i,\;\forall i\in J \\
\xi _i &\mbox{\rm{$:=$}}&\omega _i\cap \omega ,\;\forall i\notin J \\
\xi _i^{\prime } &\mbox{\rm{$:=$}}&\omega _i\cap \omega ^{\prime },\;\forall
i\notin J \\
\eta _0 &\mbox{\rm{$:=$}}&\omega \backslash (\sqcup _{i\in J}\omega _i),
\end{eqnarray*}
where $l\in \{0,\ldots ,n-1\}$; $(\eta _0,\ldots ,\eta _l)\in \frak{P}%
_\emptyset ^{l+1}(\omega )$; $(\xi _{l+1},\ldots \xi _n)\in \frak{P}%
_\emptyset ^{n-l}(\eta _0)$; $(\xi _{l+1}^{\prime },\ldots \xi _n^{\prime
})\in \frak{P}^{n-l}(\omega ^{\prime })$. This implies that (\ref{6eq68})
can be rewritten as 
\begin{eqnarray}
&&\int_{\Omega _{\Lambda \backslash \Lambda ^{\prime }}}\psi (\omega \cup
\omega ^{\prime })\nu _{z\sigma ^\tau }(d\omega )  \label{6eq69} \\
&+&\sum_{n=2}^\infty \sum_{l=0}^{n-1}\frac 1{l!(n-l)!}\int_{\Omega _{\Lambda
\backslash \Lambda ^{\prime }}}\sum_{(\eta _0,\ldots ,\eta _l)\in \frak{P}%
_\emptyset ^{l+1}(\omega )}\prod_{i=1}^l\psi (\eta _i)\varphi _{n,l}(\eta
_0)\nu _{z\sigma ^\tau }(d\omega ),  \nonumber
\end{eqnarray}
where 
\[
\varphi _{n,l}(\eta _0):=\sum_{(\xi )_{l+1}^n\in \frak{P}_\emptyset
^{n-l}(\eta _0)}\sum_{(\xi ^{\prime })_{l+1}^n\in \frak{P}^{n-l}(\omega
^{\prime })}\prod_{i=l+1}^n\psi (\xi _i^{\prime }\cup \xi _i). 
\]
Then using Lemma \ref{6eq38} we obtain 
\begin{equation}
\sum_{n=2}^\infty \sum_{l=0}^{n-1}\frac 1{l!(n-l)!}\left( \int_{\Omega
_{\Lambda \backslash \Lambda ^{\prime }}}\psi (\omega )\nu _{z\sigma ^\tau
}(d\omega )\right) ^l\int_{\Omega _{\Lambda \backslash \Lambda ^{\prime
}}}\varphi _{n,l}(\eta _0)\nu _{z\sigma ^\tau }(d\eta _0).  \label{6eq133}
\end{equation}

\noindent First we look at the integral of $\varphi _{n,l}$, 
\begin{eqnarray*}
&&\int_{\Omega _{\Lambda \backslash \Lambda ^{\prime }}}\sum_{(\xi
)_{l+1}^n\in \frak{P}_\emptyset ^{n-l}(\eta _0)}\sum_{(\xi ^{\prime
})_{l+1}^n\in \frak{P}^{n-l}(\omega ^{\prime })}\prod_{i=l+1}^n\psi (\xi
_i^{\prime }\cup \xi _i)\nu _{z\sigma ^\tau }(d\eta _0) \\
&=&\sum_{(\xi ^{\prime })_{l+1}^n\in \frak{P}^{n-l}(\omega ^{\prime
})}\int_{\Omega _{\Lambda \backslash \Lambda ^{\prime }}}\sum_{(\xi
)_{l+1}^n\in \frak{P}_\emptyset ^{n-l}(\eta _0)}\prod_{i=l+1}^n\psi (\xi
_i^{\prime }\cup \xi _i)\nu _{z\sigma ^\tau }(d\eta _0).
\end{eqnarray*}
Once more we apply Lemma \ref{6eq38} to the right hand side of the above
equality to get 
\begin{equation}
\sum_{(\xi ^{\prime })_{l+1}^n\in \frak{P}^{n-l}(\omega ^{\prime
})}\prod_{i=l+1}^n\int_{\Omega _{\Lambda \backslash \Lambda ^{\prime }}}\psi
(\xi _i^{\prime }\cup \xi _i)\nu _{z\sigma ^\tau }(d\xi _i).  \label{6eq70}
\end{equation}
Hence interchanging the sums and putting together (\ref{6eq133}) and (\ref
{6eq70}) we get 
\begin{eqnarray*}
&&\sum_{l=0}^\infty \sum_{n=l+1}^\infty \frac 1{l!}\left( \int_{\Omega
_{\Lambda \backslash \Lambda ^{\prime }}}\!\!\psi (\omega )\nu _{z\sigma
^\tau }(d\omega )\right) ^l \\
&&\times \frac 1{(n-l)!}\sum_{(\xi ^{\prime })_1^{n-l}\in \frak{P}%
^{n-l}(\omega ^{\prime })}\prod_{i=1}^{n-l}\int_{\Omega _{\Lambda \backslash
\Lambda ^{\prime }}}\psi (\omega \cup \xi _i^{\prime })\nu _{z\sigma ^\tau
}(d\omega ) \\
&=&\exp \!\left( \int_{\Omega _{\Lambda \backslash \Lambda ^{\prime
}}}\!\!\!\!\!\psi (\omega )\nu _{z\sigma ^\tau }(d\omega )\right) \exp
^{*}\!\!\left( \int_{\Omega _{\Lambda \backslash \Lambda ^{\prime
}}}\!\!\!\!\!1\!\!1_{\Omega _{fin}\backslash \{\emptyset \}}(\cdot )\psi
(\cdot \cup \omega )\nu _{z\sigma ^\tau }(d\omega )\right) (\omega ^{\prime
}).
\end{eqnarray*}
\hfill $\blacksquare $

\subsection{Proof of Proposition \ref{6eq12}\label{6eq72}}

\noindent \textbf{Proposition \ref{6eq12}} Let $\omega ,\zeta \in \Omega
_{fin}$ with $\gamma _\omega \cap \gamma _\zeta =\emptyset $. The solution
of (\ref{6eq54}) for $\omega =\{\hat{x}_1,\ldots ,\hat{x}_l\}$, $l\geq 1$
has the form 
\begin{equation}
Q(\{\hat{x}_1,\ldots ,\hat{x}_l\},\zeta )=\sum_{(\omega _1,\ldots ,\omega
_l)\in \frak{P}_\emptyset ^l(\zeta )}Q(\{\hat{x}_1\},\omega _1)\cdots Q(\{%
\hat{x}_l\},\omega _l),  \label{6eqA84}
\end{equation}
where 
\begin{equation}
Q(\{\hat{x}\},\zeta ):=(e^{2\beta B})^{|\zeta |+1}\sum_{T\in \frak{T}(\{\hat{%
x}\}\cup \zeta )}\prod_{\{\hat{y},\hat{y}^{\prime }\}\in T}|e^{-\beta \phi (%
\hat{y},\hat{y}^{\prime })}-1|,  \label{6eqA85}
\end{equation}
for $\zeta \neq \emptyset $ and $Q(\{\hat{x}\},\emptyset ):=e^{2\beta B}$.
In the case $\omega =\emptyset $ we define $Q(\emptyset ,\zeta )$ as in (\ref
{6eq55}).

\noindent \textbf{Proof.} For $\omega =\emptyset $ the assertion follows by
definition, hence we assume $\omega \neq \emptyset $. We prove the result by
induction in $|\omega |+|\zeta |$.

For $|\omega |+|\zeta |=1$ with $\gamma _\omega \cap \gamma _\zeta
=\emptyset $ we have $\zeta =\emptyset $ and $\omega =\{\hat{x}\}$. On the
one hand the r.h.s.~of (\ref{6eq54}) yields 
\[
e^{2\beta B}\sum_{\hat{\omega}\subset \emptyset }Q_I(\hat{\omega},\emptyset
\backslash \hat{\omega})|k_{\hat{\omega}}(\hat{x})|=e^{2\beta
B}Q_I(\emptyset ,\emptyset )=e^{2\beta B}, 
\]
on the other hand equation (\ref{6eqA85}) gives 
\[
Q_I(\{\hat{x}\},\emptyset )=e^{2\beta B}\sum_{T\in \frak{T}(\{\hat{x}%
\})}\prod_{\{\hat{x},\hat{x}^{\prime }\}\in \emptyset }|e^{-\beta \Phi (\hat{%
x},\hat{x}^{\prime })}-1|=e^{2\beta B}. 
\]
Thus the initial induction step is verified.

Let us assume that the result is true for $|\omega |+|\zeta |=n-1$ with $%
\gamma _\omega \cap \gamma _\zeta =\emptyset $. Choose $\omega ,\zeta $ such
that $\omega \neq \emptyset $, $\gamma _\omega \cap \gamma _\zeta =\emptyset 
$, $|\omega |+|\zeta |=n$, and denote $I(\omega )=x_0\in \omega $. Using (%
\ref{6eqA84}) for $n-1$ in the r.h.s.~of (\ref{6eq54}) one obtains 
\begin{equation}
e^{2\beta B}\sum_{\hat{\omega}\subset \zeta }|k_{\hat{\omega}}(\hat{x}%
_0)|\sum_{(\omega _{\hat{x}}^{\prime })_{\hat{x}\in \omega \backslash \{\hat{%
x}_0\}\cup \hat{\omega}}\in \frak{P}_\emptyset ^{n^{\prime }}(\zeta
\backslash \hat{\omega})}\prod_{\hat{x}\in \omega \backslash \{\hat{x}%
_0\}}Q(\{\hat{x}\},\omega _{\hat{x}}^{\prime })\prod_{\hat{x}\in \hat{\omega}%
}Q(\{\hat{x}\},\omega _{\hat{x}}^{\prime }),  \label{6eq11}
\end{equation}
where $n^{\prime }=|\omega |+|\hat{\omega}|-1.$ If $\hat{\omega}\neq
\emptyset $, then define $\omega _{\hat{x}_0}^{\prime }:=\hat{\omega}\sqcup
\bigsqcup_{\hat{x}\in \hat{\omega}}\omega _{\hat{x}}^{\prime }$ and we make
the following re-arrangement in one to one form 
\[
\emptyset \neq \hat{\omega}\subset \zeta ,(\omega _{\hat{x}}^{\prime })_{%
\hat{x}\in \omega \backslash \{\hat{x}_0\}\cup \hat{\omega}}\in \frak{P}%
_\emptyset ^{|\omega |+|\hat{\omega}|-1}(\zeta \backslash \hat{\omega}) 
\]
\[
\updownarrow 
\]
\[
((\omega _{\hat{x}}^{\prime })_{\hat{x}\in \omega },\hat{\omega},(\omega _{%
\hat{x}}^{\prime })_{\hat{x}\in \hat{\omega}}), 
\]
where $(\omega _{\hat{x}}^{\prime })_{\hat{x}\in \omega }\in \frak{P}%
_\emptyset ^{|\omega |}(\zeta )$, $\omega _{\hat{x}_0}^{\prime }\neq
\emptyset $, $\emptyset \neq \hat{\omega}\subset \omega _{\hat{x}_0}^{\prime
}$, and $(\omega _{\hat{x}}^{\prime })_{\hat{x}\in \hat{\omega}}\in \frak{P}%
_\emptyset ^{|\hat{\omega}|}(\omega _{\hat{x}_0}^{\prime }\backslash \hat{%
\omega})$. With this, the expression in (\ref{6eq11}) can be rewritten as 
\begin{eqnarray}
&&\sum_{\QATOP{(\omega _{\hat{x}}^{\prime })_{\hat{x}\in \omega }\in \frak{P}%
_\emptyset ^{|\omega |}(\zeta )}{\omega _{\hat{x}_0}^{\prime }\neq \emptyset 
}}\prod_{\hat{x}\in \omega \backslash \{\hat{x}_0\}}Q(\{\hat{x}\},\omega _{%
\hat{x}}^{\prime })\sum_{\emptyset \neq \hat{\omega}\subset \omega _{\hat{x}%
_0}^{\prime }}e^{2\beta B}|k_{\hat{\omega}}(\hat{x}_0)|  \label{6eq14} \\
&&\times \sum_{(\omega _{\hat{x}}^{\prime })_{\hat{x}\in \hat{\omega}}\in 
\frak{P}_\emptyset ^{|\hat{\omega}|}(\omega _{\hat{x}_0}^{\prime }\backslash 
\hat{\omega})}\prod_{\hat{x}\in \hat{\omega}}Q(\{\hat{x}\},\omega _{\hat{x}%
}^{\prime }).  \nonumber
\end{eqnarray}
Next we use the explicit form for $Q(\{\hat{x}\},\omega _{\hat{x}}^{\prime
}) $ in (\ref{6eqA85}) to write the term 
\[
e^{2\beta B}\sum_{\emptyset \neq \hat{\omega}\subset \omega _{\hat{x}%
_0}^{\prime }}|k_{\hat{\omega}}(\hat{x}_0)|\sum_{(\omega _{\hat{x}}^{\prime
})_{\hat{x}\in \hat{\omega}}\in \frak{P}_\emptyset ^{|\hat{\omega}|}(\omega
_{\hat{x}_0}^{\prime }\backslash \hat{\omega})}\prod_{\hat{x}\in \hat{\omega}%
}Q(\{\hat{x}\},\omega _{\hat{x}}^{\prime }) 
\]
as 
\begin{eqnarray}
\hspace{-1cm} &&\sum_{\emptyset \neq \hat{\omega}\subset \omega _{\hat{x}%
_0}^{\prime }}e^{2\beta B}\sum_{(\omega _{\hat{x}}^{\prime })_{\hat{x}\in 
\hat{\omega}}\in \frak{P}_\emptyset ^{|\hat{\omega}|}(\omega _{\hat{x}%
_0}^{\prime }\backslash \hat{\omega})}\prod_{\hat{x}\in \hat{\omega}%
}(e^{2\beta B})^{|\omega _{\hat{x}}^{\prime }|+1}\sum_{T_{\hat{x}}\in \frak{T%
}(\{\hat{x}\}\cup \omega _{\hat{x}}^{\prime })}\!\!\!|k_{T_{\hat{x}%
}}|\,\,|k_{\hat{\omega}}(\hat{x}_0)|  \nonumber \\
\hspace{-1cm} &=&\sum_{\emptyset \neq \hat{\omega}\subset \omega _{\hat{x}%
_0}^{\prime }}(e^{2\beta B})^{1+|\omega _{\hat{x}_0}^{\prime }\backslash 
\hat{\omega}|+|\hat{\omega}|}\!\!\!\!\!\!\!\!\!\!\sum_{(\omega _{\hat{x}%
}^{\prime })_{\hat{x}\in \hat{\omega}}\in \frak{P}_\emptyset ^{|\hat{\omega}%
|}(\omega _{\hat{x}_0}^{\prime }\backslash \hat{\omega})}\prod_{\hat{x}\in 
\hat{\omega}}\sum_{T_{\hat{x}}\in \frak{T}(\{\hat{x}\}\cup \omega _{\hat{x}%
}^{\prime })}\!\!\!\!\!\!\!|k_{T_{\hat{x}}}|\,\,|k_{\hat{\omega}}(\hat{x}%
_0)|.\quad  \label{6eq13}
\end{eqnarray}
We again make a re-arrangement: for $\emptyset \neq \hat{\omega}\subset
\omega _{\hat{x}_0}^{\prime }$, $(\omega _{\hat{x}}^{\prime })_{\hat{x}\in 
\hat{\omega}}\in \frak{P}_\emptyset ^{|\hat{\omega}|}(\omega _{\hat{x}%
_0}^{\prime }\backslash \hat{\omega})$, and $(T_{\hat{x}})_{x\in \hat{\omega}%
}\in \times _{\hat{x}\in \hat{\omega}}\frak{T}(\{\hat{x}\}\cup \omega _{\hat{%
x}}^{\prime })$ we define 
\[
T:=\bigsqcup_{\hat{x}\in \hat{\omega}}T_{\hat{x}}\sqcup \{(\hat{x},\hat{x}%
_0)|\hat{x}\in \hat{\omega}\}\in \frak{T}(\omega _{\hat{x}_0}^{\prime }\cup
\{\hat{x}_0\}), 
\]
and vice versa, given $T\in \frak{T}(\omega _{\hat{x}_0}^{\prime }\cup \{%
\hat{x}_0\})$ we define $\hat{\omega}$, $\omega _{\hat{x}}^{\prime }$, and $%
(T_{\hat{x}})_{\hat{x}\in \hat{\omega}}$ by 
\begin{eqnarray*}
\hat{\omega} &\mbox{\rm{$:=$}}&\{\hat{x}\in V(T)|(\hat{x},\hat{x}_0)\in
T\}\subset \omega _{\hat{x}_0}^{\prime } \\
T_{\hat{x}_0}\oplus \bigoplus_{\hat{x}\in \hat{\omega}}T_{\hat{x}} &%
\mbox{\rm{$:=$}}&T\backslash \{(\hat{x},\hat{x}_0)|\hat{x}\in \hat{\omega}%
\}\;\mathrm{with}\;\hat{x}\in V(T_{\hat{x}})\;\mathrm{and}\;V(T_{\hat{x}_0})=%
\hat{x}_0 \\
\omega _{\hat{x}}^{\prime } &\mbox{\rm{$:=$}}&V(T_{\hat{x}})\backslash \{%
\hat{x}\}
\end{eqnarray*}

Then (\ref{6eq13}) can be written (using \ref{6eqA85}) as 
\[
(e^{2\beta B})^{1+|\omega _{\hat{x}_0}^{\prime }|}\sum_{T\in \frak{T}(\omega
_{\hat{x}_0}^{\prime }\cup \{\hat{x}_0\})}k_T=Q(\{\hat{x}_0\},\omega _{\hat{x%
}_0}^{\prime }). 
\]
Hence (\ref{6eq14}) now simplifies to 
\[
\sum_{\QATOP{(\omega _{\hat{x}}^{\prime })_{\hat{x}\in \omega }\in \frak{P}%
_\emptyset ^{|\omega |}(\zeta )}{\omega _{\hat{x}_0}^{\prime }\neq \emptyset 
}}\prod_{\hat{x}\in \omega \backslash \{\hat{x}_0\}}Q(\{\hat{x}\},\omega _{%
\hat{x}}^{\prime })Q(\{\hat{x}_0\},\omega _{\hat{x}_0}^{\prime }). 
\]
After an explicit calculation for the case $\hat{\omega}=\emptyset $ we see
that the above expression is nothing but the required form for $Q(\omega
,\zeta )$.\hfill $\blacksquare $

\subsection{Proof of Proposition \ref{6eq74} \label{6eq75}}

\noindent \textbf{Lemma \ref{6eq128} }For every $\hat{x}\in X\times S$, $%
Y\in \mathcal{B}(X)$, and $n\geq 1$ we have 
\begin{eqnarray}
&&\int_{(Y\times S)^n}Q(\{\hat{x}\},\{\hat{y}\}_1^n)\sigma ^\tau (d\hat{y}%
)_1^n  \label{6eq86} \\
&\leq &e^{2\beta B(n+1)}C(\beta )^{n-1}(n+1)^{n-1}\int_{Y\times S}|e^{-\beta
\phi (\hat{x},\hat{y})}-1|\sigma ^\tau (d\hat{y}).  \nonumber
\end{eqnarray}

\noindent \textbf{Proof.} In the following we denote $\hat{y}_{n+1}:=\hat{x}$%
. The equality (\ref{6eq10}) implies the following estimate for (\ref{6eq86}%
) 
\begin{equation}
(e^{2\beta B})^{n+1}\sum_{T\in \frak{T}([n+1])}\int_{(Y\times
S)^n}\prod_{(i,j)\in T}|e^{-\beta \Phi (\hat{y}_i,\hat{y}_j)}-1|\sigma ^\tau
(d\hat{y})_1^n.  \label{6eq134}
\end{equation}
We now estimate by induction in $n$ the term 
\begin{equation}
\int_{(Y\times S)^n}\prod_{(i,j)\in T}|e^{\beta \Phi (\hat{y}_i,\hat{y}%
_j)}-1|\sigma ^\tau (d\hat{y})_1^n.  \label{6eq71}
\end{equation}
For $n=1$ all trees $T$ are of the form $\{\{\hat{x},\hat{y}_1\}\}$ and
hence (\ref{6eq71}) is reduced to 
\[
\int_{(Y\times S)}|e^{\beta \Phi (\hat{x},\hat{y}_1)}-1|\sigma ^\tau (d\hat{y%
}_1).
\]

\noindent Let us assume that for $n=N-1$ we have for all $T\in \frak{T}%
([n+1])$%
\begin{eqnarray*}
&&\int_{(Y\times S)^n}\prod_{(i,j)\in T}|e^{-\beta \Phi (\hat{y}_i,\hat{y}%
_j)}-1|\sigma ^\tau (d\hat{y})_1^n \\
&\leq &C(\beta )^{n-1}\int_{Y\times S}|e^{-\beta \phi (\hat{y}_{n+1},\hat{y}%
)}-1|\sigma ^\tau (d\hat{y}).
\end{eqnarray*}
For the case $n=N$ we proceed as follows. Let $T\in \frak{T}([n+1])$ be
given. Choose $\hat{y}_{n+1}$ as a foot point of $T$. Then there exists a
final pair $\{j_1,j_2\}\in T$ where $\hat{y}_{j_1}$ is the final vertex and $%
\hat{y}_{j_1}\neq \hat{y}_{n+1}$. This implies the following estimate 
\begin{eqnarray*}
&&\int_{(Y\times S)^n}\prod_{(i,j)\in T}|e^{-\beta \phi (\hat{y}_i,\hat{y}%
_j)}-1|\sigma ^\tau (d\hat{y})_1^n \\
&\leq &\int_{(Y\times S)^{n-1}}\prod_{(i,j)\in T\backslash
\{j_1,j_2\}}|e^{-\beta \phi (\hat{y}_i,\hat{y}_j)}-1| \\
&&\hspace{2cm}\times \int_{Y\times S}|e^{-\beta \phi (\hat{y}_{j_1},\hat{y}%
_{j_2})}-1|\sigma ^\tau (d\hat{y}_{j_1})\prod_{\QATOP{l=1}{l\neq j_1}%
}^n\sigma ^\tau (d\hat{y}_l) \\
&\leq &C(\beta )\int_{(Y\times S)^{n-1}}\prod_{(i,j)\in T\backslash
\{j_1,j_2\}}|e^{-\beta \phi (\hat{y}_i,\hat{y}_j)}-1|\prod_{\QATOP{l=1}{%
l\neq j_1}}^n\sigma ^\tau (d\hat{y}_l) \\
&\leq &C(\beta )^{n-1}\int_{Y\times S}|e^{-\beta \phi (\hat{y}_{n+1},\hat{y}%
)}-1|\sigma ^\tau (d\hat{y}).
\end{eqnarray*}
where in the last inequality we used the induction step. Thus (\ref{6eq134})
yields
\[
(e^{2\beta B})^{n+1}C(\beta )^{n-1}\int_{Y\times S}|e^{-\beta \phi (\hat{y}%
_{n+1},\hat{y})}-1|\sigma ^\tau (d\hat{y})\sum_{T\in \frak{T}([n+1])}1.
\]
It follows from Proposition~\ref{6eq110} that $|\frak{T}([n+1])|=(n+1)^{n-1}$%
.\hfill $\blacksquare $

\end{appendix}

\addcontentsline{toc}{section}{References}

\newcommand{\etalchar}[1]{$^{#1}$}

\end{document}